\documentclass[aps,prd,superscriptaddress,twocolumn]{revtex4-2}
\usepackage{blindtext}

\usepackage{mathtools,mathalfa,amssymb,amsmath,amsfonts,amsthm,bm}
\usepackage[T1]{fontenc}
\usepackage{tipa}
\usepackage{wasysym}
\usepackage{appendix}
\usepackage{calrsfs}
\usepackage{tikz}
\usepackage[font=small,
   justification=justified,
   format=plain]{caption}
\usepackage{aas_macros}
\usetikzlibrary {decorations.pathmorphing}
\usepackage{orcidlink}

\usepackage{soul}

\usepackage{framed}
\usepackage{color}
\usepackage[mathscr]{euscript}

\renewcommand{\d}{\mathrm{d}}

\newcounter{definition}
\definecolor{wuppergreen}{RGB}{137, 186, 23}

\newcommand{\pullback}[1]{\hbox{\lower0.5ex\hbox{${}_{\leftarrow}$}}\kern-1.9ex{#1}}
\newcommand{\pullbacklong}[1]{\hbox{\lower0.85ex\hbox{${}_{\longleftarrow}$}}\kern-3.0ex{#1}}
\newcommand{\pullbackllong}[1]{\hbox{\lower0.85ex\hbox{${}_{\longleftarrow\!\!-\!\!-\!\!-\!\!-}$}}\kern-6.4ex{#1}}

\usepackage{graphicx}
\usepackage{subcaption}
\graphicspath{{./figures-paper/}}

\begin{document}
\title{Wave optics for rotating stars}

\author{B\'eatrice Bonga
\orcidlink{0000-0002-5808-9517}
}
\email{bbonga@science.ru.nl}
\affiliation{Institute for Mathematics, Astrophysics and Particle Physics, Radboud University, Heyendaalseweg 135, 6525 AJ Nijmegen, The Netherlands}

\author{Job Feldbrugge
\orcidlink{0000-0003-2414-8707}}

\email{job.feldbrugge@ed.ac.uk}
\affiliation{Higgs Centre for Theoretical Physics, University of Edinburgh, James Clerk Maxwell Building, Edinburgh EH9 3FD, UK}

\author{Ariadna Ribes Metidieri
\orcidlink{0009-0001-4084-2259}}

\email{ariadna.ribesmetidieri@ru.nl}
\affiliation{Institute for Mathematics, Astrophysics and Particle Physics, Radboud University, Heyendaalseweg 135, 6525 AJ Nijmegen, The Netherlands}

\begin{abstract}
  Gravitational lensing in wave optics is a rich field combining caustic singularities, general relativity and interference phenomena. We present a detailed evaluation of wave optics effects resulting from the frame-dragging of a rotating star modeled by a Lense-Thirring spacetime. We demonstrate that, contrary to what was previously stated in the literature (see e.g. Ref.~\cite{Baraldo:1999}), the spin of the star leaves an intricate imprint on the interference fringes and the caustics of the lensed source. This interference pattern can in principle be used to directly measure the spin of the lens.
\end{abstract}

\maketitle

\section{Introduction}
Einstein's theory of gravity has reshaped our understanding of the Universe, both at large scales and about the objects within it, ranging from black holes and gravitational waves to gravitational lensing. Each phenomenon is of great interest and can potentially reveal new physics. In this paper, we reconsider gravitational lensing in the wave optics regime by a rotating star. 

Over the last decades, tremendous progress has been made in the theory and observation of gravitational lensing in astronomy using the geometric optics approximation, treating light as rays following null geodesics in curved spacetime. The geometric approximation suffices for most astrophysical settings. 
However, wave optics are important when you have coherent (nearly) monochromatic waves.
In such cases, as infinitely many rays with nearly equal path lengths converge at the observer, the resulting phenomena of diffraction and interference invalidate geometrical optics. 
With the advent of numerous Fast Radio Bursts and pulsar observations \cite{Petroff:2019}, and the first gravitational waves detections \cite{Abott:2016}, all emitting coherent long wavelength radiation, it may become possible to observe such interference effects due to gravitational lensing for the first time. This is particularly important in the vicinity of caustics, where the intensity is amplified and the geometric approximation breaks down. 

The wave nature of radiation in gravitational lensing was first formulated by \cite{Nakamura:1999, Baraldo:1999}. Recently, much progress has been made on the numerical evaluation of the interference patterns resulting from lensing in wave optics using both Fourier \cite{Grillo:2018}, and more general complex analysis methods \cite{Feldbrugge:2020, Tambalo:2023, Feldbrugge:2023,Feldbrugge:2023b, Jow:2023}. For a path integral analysis of gravitational lensing in wave optics see \cite{Braga:2024}.

The effect of rotation of the lens is clearly also important. This is already evident in the geometrics optics limit. 
As the star rotates, radiation traveling in the direction of rotation of the object will move past the massive object faster than radiation moving against the rotation, as seen by a distant observer. This is known as frame-dragging or the Lense-Thirring effect \cite{Thirring:1918, Lense:1918}. Relativistic frame-dragging was recently observed in relativistic jets \cite{Miller-Jones:2019} and in a pulsar white dwarf binary system \cite{Venkatraman:2020}. Moreover, in the coming years, the Lense-Thirring effect on the orbits of the S2 star around the supermassive black hole in the center of our Milky Way may be detected with the GRAVITY instrument of the Very Large Telescope \cite{Grould:2017}.

The simplest situation in which wave optics are important is the case in which the lens  is modelled as a point mass and the source, lens and observer are all far away from each other, so that the thin lens approximation applies. This case has been extensively studied in the weak field regime in which (post-)Newtonian theory applies. Remarkably, the intensity pattern in this case can be evaluated completely analytically \cite{Nakamura:1999}. Generalizations of this scenario include a singular isothermal sphere lens \cite{Ulmer:1994ij,Takahashi:2003ix,Matsunaga:2006uc} and a replacement of a single point mass by a binary \cite{Feldbrugge:2020}. 

In this article, we investigate the effect of a rotating lens. We account for the spin by using the Lens-Thirring metric, \textit{i.e.}, the slow-spin approximation of the Kerr metric. This was previously studied in \cite{Baraldo:1999}, in which the authors concluded that the interference pattern of a rotating star has the same \emph{shape} as for the non-rotating lens, but that the wave pattern is shifted \emph{translationally} in the direction perpendicular to the angular momentum vector projected onto the lens plane. Other papers using similar arguments also concluded that the interference pattern of the rotating lens is degenerate with that of the non-rotating lens up to this translational shift \cite{Asada:2000vn,Sereno:2002tv,Ebrahimnejad:2005}. As a result, it was argued that a rotating lens is not distinguishable from a non-rotating one (unless one has an independent way of knowing the precise location of the source). 
We find that this is in fact incorrect: rotation of the lensing object does change the interference pattern beyond simply shifting it. Earlier results were based on a seemingly innocent mathematical transformation that is in fact highly singular. By not using this transformation, we find that the phenomenology of the rotating lens is much richer than merely shifting the interference pattern. Hence, in principle, the rotating lens is not degenerate with the non-rotating one and one can determine how fast the lens is rotating based on its interference pattern.

The outline of this paper is as follows. In Sec.~\ref{sec:rotating-stars}, we introduce the setup and describe our methods for evaluating the interference pattern. The results are discussed in Sec.~\ref{sec:results}, where we show explicit interference patterns and contrast these with the shifted interference pattern as argued for in \cite{Baraldo:1999, Asada:2000vn,Sereno:2002tv,Ebrahimnejad:2005}. In App.~\ref{ap:unfolding}, we demonstrate explicitly how sensitive critical and caustic curves are to small perturbations. App.~\ref{sec:small-spin} discusses the small spin approximation and some of its limitations.

\section{Rotating stars}\label{sec:rotating-stars}

In this section, we provide a lightning review of the key quantity relevant for wave optics: the Kirchhoff-Fresnel integral with its integrand determined by the time-delay function. Next, we sketch the derivation of the time delay function for a rotating star, evaluate the caustics, and propose a method to efficiently evaluate the resulting Kirchhoff-Fresnel integral.

In the presence of a lens, the wave amplitude of a point source assumes the form of a Kirchhoff-Fresnel integral
\begin{align}
\label{eq:KFintegral}
  \Psi(\bm{y}) = \left(\frac{w}{2\pi i}\right)^{d/2} \int_L e^{i w T(\bm{x},\bm{y})}\mathrm{d}\bm{x}\,,
\end{align}
with the angular frequency of the radiation $w$, the point $\bm{x}$ on the $d$-dimensional lens plane, the relative position of the observer and source $\bm{y}$ in the image plane and the time delay function 
\begin{align}
  T(\bm{x},\bm{y}) = \frac{(\bm{x}-\bm{y})^2}{2} - \varphi(\bm{x})\,,
\end{align}
governed by a geometric term $(\bm{x}-\bm{y})^2/2$ and the phase variation $\varphi$ modeling the effect of the lens. The integral in Eq.~\eqref{eq:KFintegral} can be interpreted as the superposition of all possible rays propagating from the source to the observer and intersecting the lens plane at $\bm{x}$ weighted by the phase $e^{i w T}$ \cite{Feynman:2006}. Note that the interference integral is given in dimensionless units, $w,\bm{x},$ and $\bm{y}$ and that the amplitude is normalized with respect to the unlensed case, \textit{i.e.}, when $\varphi(\bm{x})=0$, the amplitude $\Psi(\bm{y})=1$. For more details, see \cite{Schneider:1992, Nakamura:1999}.

\subsection{The time delay}
\label{sec:time-delay}
The time delay function for a rotating star (see Eq.~\ref{eq:time-delay-dimensionless}) has appeared in the literature before \cite{Ibanez:1983, Dymnikova:1986, Glicenstein:1999, Baraldo:1999}, here we review its derivation. A busy reader may skip this subsection. 

Gravitational lensing is governed by the time as measured by the observer that light rays take to travel from the source to the observer while following geodesics in the curved spacetime. The source is modelled by a point source that emits spherical, monochromatic waves with frequency $\omega$. The emitted light then propagates past a rotating gravitational lens before reaching a distant observer. The space-time metric of a rotating star in the weak field limit is
\begin{align}
  \mathrm{d}s^2 = &-\left(1 + 2 U(\bm{r})\right) c^2 \mathrm{d}t^2 + (1-2U(\bm{r}))\mathrm{d}\bm{r}^2 \nonumber\\
  &-\frac{4 G \varepsilon_{j k m} J^k x^m \mathrm{d} x^j \mathrm{d} t}{c^2 \lVert \bm{r} - \bm{r}^* \rVert^3} \, ,\label{eq:metric}
\end{align}
which can be interpreted as the Minkowski metric, plus a small disturbance due to the gravitational potential $U(\bm{r}) = - GM /(c^2\lVert \bm{r} - \bm{r}^*\rVert)$ of the star of mass $M$ located at $\bm{r}^*$, and a frame-dragging term resulting from the spin $\bm{J}$ of the lens. 

When the particle’s deflection occurs over a relatively small region, relative to the distance from the source to the observer, one can use the thin lens approximation. 
In this approximation, the time delay function consists of three components modulo some constant terms that do not affect the wave amplitude. First, the distance of the lensed path exceeds that of a straight line from the source to the observer by $\frac{1}{2} \frac{D_{OL}D_{OS}}{D_{OS} - D_{OL}} \lVert \bm{\theta} - \bm{\theta}_{S}\rVert^2$ with the angular diameter distances from the observer to the lens $D_{OL}$ and the observer to the source $D_{OS}$ respectively. $\bm{\theta}_{S}$ stands for the angular position of the source as measured from the line of sight, and $\bm{\theta}$ is the angular position of the ray crossing the lens plane $\bm{\theta}$ with respect to the position of the lens on the sky (see Fig.~\ref{fig:setup-paper} for a comprehensive schematic of the setup). 
Second, the line integral of the potential $\int U(\bm{r})\mathrm{d}r$ for small angles assumes the form $4GM/c^2 \log \lVert \bm{\theta} - \bm{\theta}^*\rVert$ up to a constant. Third, the frame-dragging term yields a speed-up/delay for passing along/opposite the spin direction of the star by
$\frac{4G }{c^3 D_{OL}} \frac{( \bm{J} \times \bm{n})\cdot \bm{\theta}}{\lVert\bm{\theta}\rVert^2}$. 
In terms of dimensionless units, we obtain the time delay function 
\begin{align}
  T(\bm{x},\bm{y}) &= \frac{(\bm{x}-\bm{y})^2}{2}  -\log x + \frac{\bm{\alpha} \cdot \bm{x}} {x^2}\,,
  \label{eq:time-delay-dimensionless}
\end{align} 
with the norm $x= \lVert \bm{x} \rVert$, the spin vector
\begin{align}
    \bm{\alpha} = \frac{ \bm{J}\times \bm{n}}{c M r_E}\,,
\end{align}
where $\bm{n}$ is a unit vector pointing along the line of sight towards the source,  and the Einstein radius 
\begin{align}
  r_E
  = \sqrt{ \frac{4 GM}{c^2} \frac{D_{OL}(D_{OS} - D_{OL})}{D_{OS}}} \,.
\end{align}
In this formula, the angles 
\begin{align}\label{eq:normalization-x-y}
  \bm{x} = \frac{\bm{\theta}}{\theta_E }\,, \quad \bm{y} =  \frac{\bm{\theta}_{S}}{\theta_E}\,,
\end{align} 
are normalized with respect to the Einstein angle 
\begin{align}
  \theta_E 
  &= r_E / D_{OL} =\sqrt{ \frac{4 GM}{c^2} \frac{D_{OS} - D_{OL}}{D_{OL}D_{OS}}}\,.
\end{align}
In addition, we work with the dimensionless angular frequency $w=4 G M \omega/c^2$, which is defined as the frequency of the radiation $\omega$ normalized using the gravitational radius of the lens. The frequency also enters in the definition of the wavelength of the radiation, defined as $\lambda = 2 \pi c/ \omega$ (see \cite{Baraldo:1999}). In astrophysics, the dimensionless angular frequency $w$ is generally a large number.

It is often convenient to express the spin vector in terms of the Kerr parameter $\bm{a} = \bm{J}/(cM)$:
\begin{align}
  \bm{\alpha} = \frac{ \bm{a}\times \bm{n}}{r_E} \, .
\end{align}
 For black holes, the Kerr parameter takes values between 0 and $M$ with $M$ corresponding to an extremal black hole. As we are modeling rotating stars instead of black holes and work with the small spin limit, the Kerr parameter should be considered small, \textit{i.e.} $a/M < 1$. This implies that for realistic settings $\alpha \ll 1$, given that in the thin lens approximation both $D_{OL}$ and $D_{OS}$ are large so that $r_E$ is large as well ($r_E \sim \sqrt{2GM D_{OL}/c^2})$. 
 While the angular momentum $\bm{J}$ is a three-vector, we will from hereon interpret the spin vector $\bm{\alpha}$ as a two-vector in the lens plane, normal to the projection of the spin vector $\bm{J}$ onto the lens plane $\bm{J} - (\bm{n} \cdot \bm{J})\bm{n}$. As a result, frame dragging only alters the time delay function when the angular momentum vector $\bm{J}$ is misaligned with the line of sight $\bm{n}$ and for a given value of $J$, the effect is maximal when the angular momentum vector lies in the lens plane. Using spherical coordinates 
\begin{align}
  \bm{J}= J ( \sin\theta_0\cos\phi_0, \sin\theta_0\sin\phi_0,\cos\theta_0),
\end{align}
with the line of sight $\bm{n}=(0,0,1)$, where $\theta_0$ is the angle between the angular momentum and the line of sight, the spin vector assumes the form 
\begin{align}
  \bm{\alpha}=\frac{a \sin\theta_0}{r_E} (-\sin\phi_0,  \cos \phi_0)\,,
\end{align}
with the norm $a = \lVert \bm{a} \rVert$. The angle $\theta_0$ determines the amplitude of the spin vector, while the angle $\phi_0$ determines its orientation.
\begin{center}
    
    \includegraphics[width=\linewidth]{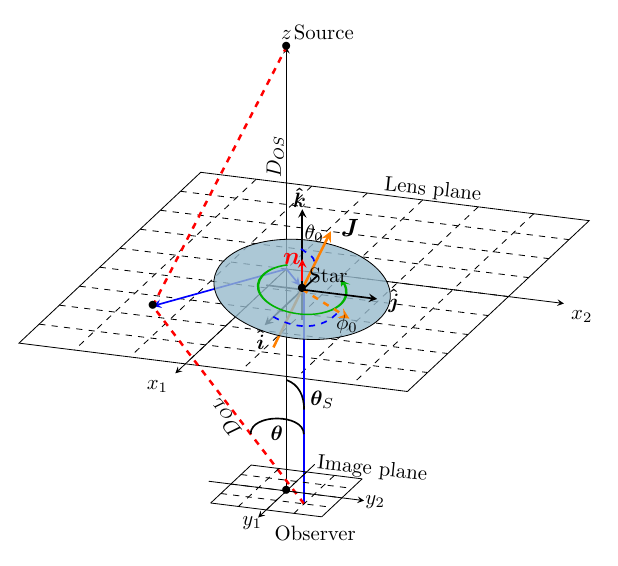}~\captionof{figure}{
       Schematic representation of the lensed null rays that arrive at the observer. In the thin lens approximation, we consider that the rays travel along straight lines (as they would in Minkowski space), and the trajectory's deviation due to the rotating star's presence only occurs in the lens plane, represented by the dashed grid. The rotation of the star drags the neighboring spacetime, which has been represented in blue.   The line connecting the source to the observer is the line of sight. We consider the star to lie at an angular distance $\bm{\theta}_S$ as measured from the line of sight. A ray intersects the lens plane at an angular distance $\bm{\theta}$ as measured from the position of the star. We define a Cartesian coordinate system centered at the star with the $z$ axis along the line of sight and pointing towards the source. Similarly, the spherical coordinate system $(\theta_0\,,\phi_0)$ is defined in the usual way. The distance separating the observer from the lens is denoted $D_{OL}$ while the distance separating the observer from the source is $D_{OS}$. Both distances are much larger than the characteristic lengths in the lens plane. In orange, we represent the angular momentum vector of the star $\bm{J}$ and in red a unit vector $\bm{n}$ parallel to the line of sight. 
       }
    \label{fig:setup-paper}
\end{center}
\bigskip
In the literature, the time delay function of a rotating star is sometimes expressed as \cite{Baraldo:1999,Asada:2000vn,Sereno:2002tv}
\begin{align}
  T(\bm{x},\bm{y}) 
  & =\frac{(\bm{x}-\bm{y})^2}{2}  -\log \lVert \bm{x} - \bm{\alpha}\rVert - \sum_{n=2}^\infty \frac{\alpha^n \cos n \bar{\theta}}{n x^n} 
  \,,\label{eq:expansion}
\end{align} 
whereby the frame dragging term $\bm{\alpha} \cdot \bm{x}/x^2$ is absorbed into the logarithm and the higher order corrections, with $\bar{\theta}$ the angle between $\bm{\alpha}$ and $\bm{x}$. These higher-order corrections, often written as $\mathcal{O}(\alpha^2)$, are then neglected in those references. To first order in the spin parameter, this time delay function is equivalent to the time delay function of the non-rotating point source with $\bm{x} \mapsto \bm{x} - \bm{\alpha}$. Based on this observation, it is then argued that the interference pattern of a rotating star is the same as the interference pattern of the non-rotating star shifted by the spin vector $\bm{\alpha}$ (see for example \cite{Baraldo:1999, Asada:2000vn,Sereno:2002tv,Ebrahimnejad:2005}). 

While this may appear completely reasonable at first, this seemingly innocent trick is in fact mathematically rather violent: The residue $\sum_{n=2}^\infty \frac{\alpha^n \cos n \bar{\theta}}{n x^n}$ diverges logarithmically at both the origin $\bm{0}$ and the point $\bm{\alpha}$ (where the sum reduces to the harmonic series $\sum_{n=2}^\infty n^{-1}$). While the first singularity can be understood as the position of the lens and also appears in the time delay function in Eq.~\eqref{eq:time-delay-dimensionless}, the latter singularity is not physical. It is merely an artifact of rewriting Eq.~\eqref{eq:time-delay-dimensionless} as  Eq.~\eqref{eq:expansion}. Since the Kirchhoff-Fresnel integral ranges over all points $\bm{x}$ in the lens plane, this way of rewriting the system is not a small perturbation of the original system, even for small $\alpha$. Moreover, it is important to realize that the rotational symmetry of the phase variation of the non-rotating lens $\varphi(\bm{x}) = \log x$ makes the interference pattern degenerate (in the sense of catastrophe theory). A small fluctuation in the phase variation $\varphi$ --- thereby breaking the rotational symmetry --- will dramatically change the caustics and interference pattern (see appendix \ref{ap:unfolding} for some concrete examples). The frame-dragging contribution certainly breaks the rotation symmetry as light rays moving along the spin direction of the star are sped up while light rays propagating opposite of the spin direction are delayed by frame dragging. By approximating the phase variation of the rotating star by 
\begin{align}
  \varphi(\bm{x}) \sim -\log \lVert \bm{x} - \bm{\alpha}\rVert\,, \label{eq:approx}
\end{align} 
the rotational symmetry and consequently the degeneracy of the non-rotating point lens is preserved. Hence, this is not a good approximation as we will also explicitly see in Sec.~\ref{sec:results}.

\subsection{Geometric optics}
Following Fermat's principle, the classical rays correspond to the stationary point of the time delay function,
\begin{align}
  \nabla_{\bm{x}} T(\bm{x},\bm{y}) = \bm{x}-\bm{y} - \frac{\bm{x} - \bm{\alpha}}{x^2} - \frac{2 (\bm{\alpha} \cdot \bm{x})\bm{x}}{x^4} = 0\,.\label{eq:variation}
\end{align}
Upon solving for $\bm{y}$, we obtain the geometric optics map, 
\begin{align}
    \bm{\xi}(\bm{x}) = \bm{x} - \frac{\bm{x} - \bm{\alpha}}{x^2} - \frac{2 (\bm{\alpha} \cdot \bm{x})\bm{x}}{x^4}\,,
\end{align}
sending points in the lens plane to the observer. In the \textit{geometric optics approximation}, the intensity assumes the form 
\begin{align}
  I_{geometric}(\bm{y}) = \sum_{\bm{x} \in \bm{\xi}^{-1}(\bm{y})} \frac{1}{|\det \nabla \bm{\xi}(\bm{x})|}\, \label{eq:geometric}
\end{align}
including a contribution for each classical ray propagating from the source to the observer passing the lens at $\bm{\xi}^{-1}(\bm{y}) = \{\bm{x}\,|\, \bm{\xi}(\bm{x}) = \bm{y}\}$ \cite{Schneider:1992}. The intensity \eqref{eq:geometric} spikes when the deformation tensor $\nabla \bm{\xi}$ is singular. In the lens plane, this corresponds to the critical curve
\begin{align}
    \mathcal{C} &= \{\bm{x}\,|\, \det \nabla \bm{\xi}(\bm{x}) =0\}\,,\\
    &= \{ \bm{x}\,|\,\bm{x}^2 ( \bm{x}^4-1) -4\, \bm{\alpha}\cdot (\bm{\alpha} + \bm{x}) = 0\}\,.
\end{align}
When mapping the critical curve to the image plane, we obtain the caustic curve,
\begin{align}
    \bm{\xi}(\mathcal{C}) = \{ \bm{\xi}(\bm{x})\,|\, \bm{x} \in \mathcal{C}\}\,,
\end{align}
consisting of the geometric pattern at which the intensity of the geometric optics approximation spikes. As we will see below, the caustics are the places where the geometric optics approximation starts to fail. Moreover, the caustics mark the locations in the image plane at which the intensity pattern in wave optics qualitatively changes. As we approach a fold/cusp caustic in the image plane the two/three rays $\bm{\xi}^{-1}(\bm{y})$ coalesce at a point on the critical curve in the lens plane. 

For the non-rotating point lens, the lens map $\bm{\xi}(\bm{x})=\bm{y}$ yields two real classical rays \cite{Nakamura:1999} 
\begin{align}
  \bm{x}_\pm = \frac{\bm{y}}{2 y}\left(y \pm \sqrt{4+y^2}\right)\,.
\end{align} 
The critical curve $\mathcal{C}$, consisting of the unit circle $\{\lVert \bm{x} \rVert = 1\}$ in the lens plane, is known as the Einstein ring. The caustic curve consist of the point $\bm{\xi}(\mathcal{C})=\{\bm{0}\}$ in the image plane. The observation that the one-dimensional unit circle is mapped to a zero-dimensional point signals that this set-up is degenerate resulting from the radial symmetry of the phase-variation $-\log x$. A small perturbation, breaking the radial symmetry of the phase variation in the lens plane dramatically changes the caustic curve into an astroid, consisting of a fold curve with four cusp points (see appendix \ref{ap:unfolding}). A similar phenomenon can be observed in the unfolding of the caustic resulting from a symmetric liquid dropped lens \cite{Nye:1978, Nye:1986}.

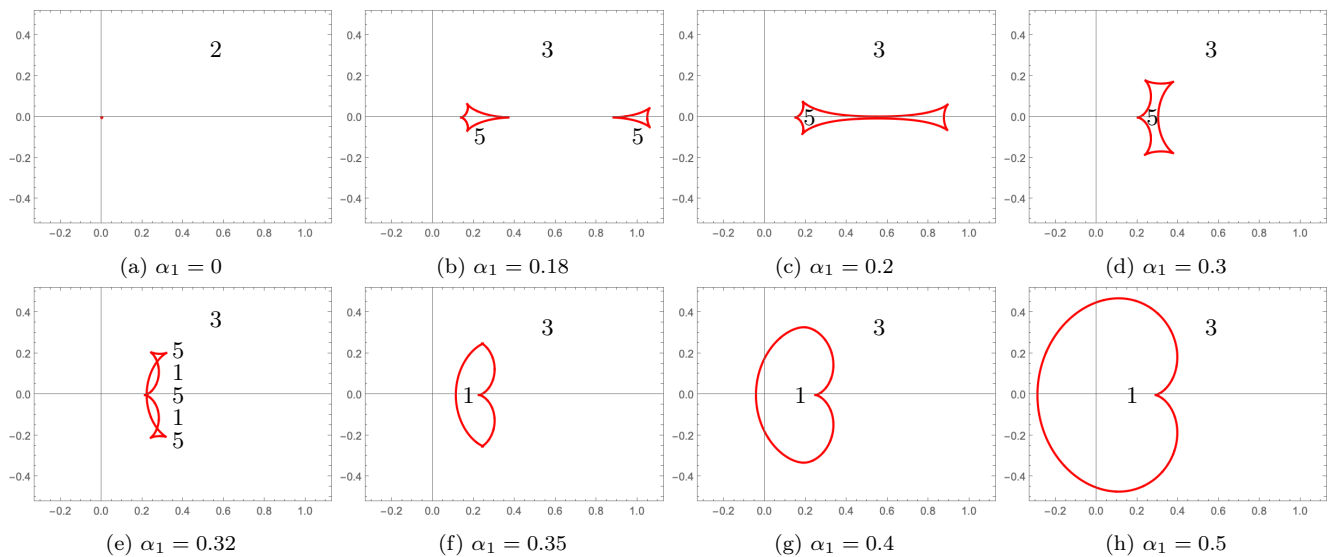
\begin{figure*}
  \begin{subfigure}[b]{0.24\linewidth}
    \begin{tikzpicture}
      \draw (0, 0) node[inner sep=0] {\includegraphics[width=\textwidth]{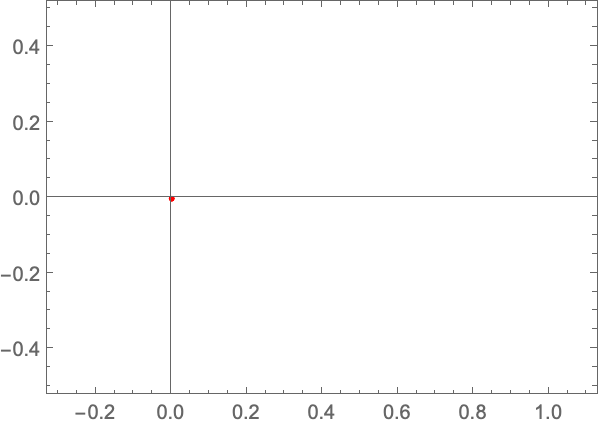}};
      \draw (0.6, 1) node {$2$};
    \end{tikzpicture}
    \caption{$\alpha_1=0$}
  \end{subfigure}
  \begin{subfigure}[b]{0.24\linewidth}
    \begin{tikzpicture}
      \draw (0, 0) node[inner sep=0] {\includegraphics[width=\textwidth]{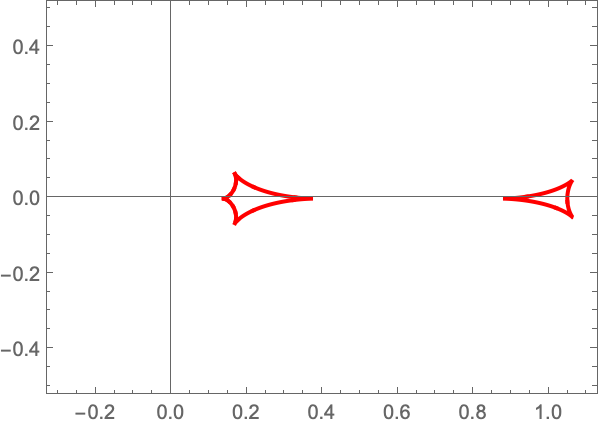}};
      \draw (0.6, 1) node {$3$};
      \draw (-0.3, -0.15) node {$5$};
      \draw (1.8, -0.15) node {$5$};
    \end{tikzpicture}
    \caption{$\alpha_1=0.18$}
   \end{subfigure}
  \begin{subfigure}[b]{0.24\linewidth}
    \begin{tikzpicture}
      \draw (0, 0) node[inner sep=0] {\includegraphics[width=\textwidth]{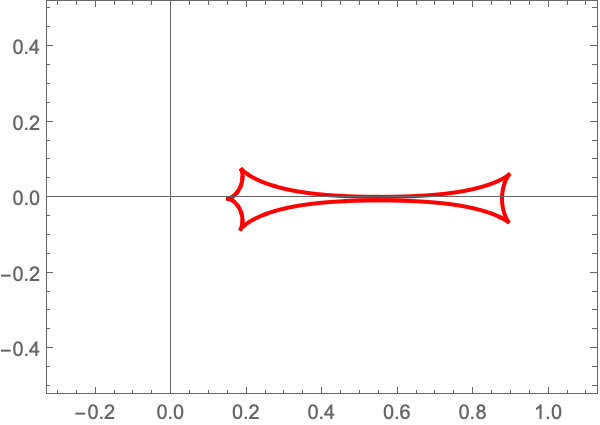}};
      \draw (0.6, 1) node {$3$};
      \draw (-0.33, 0.1) node {$5$};
    \end{tikzpicture}
    \caption{$\alpha_1=0.2$}
   \end{subfigure}
  \begin{subfigure}[b]{0.24\linewidth}
    \begin{tikzpicture}
      \draw (0, 0) node[inner sep=0] {\includegraphics[width=\textwidth]{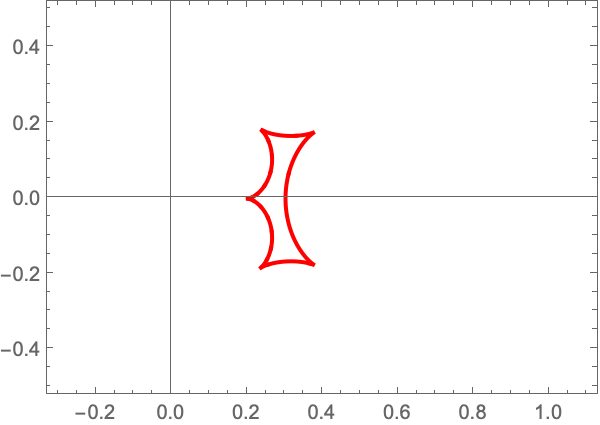}};
      \draw (0.6, 1) node {$3$};
      \draw (-0.18, 0.1) node {$5$};
    \end{tikzpicture}
    \caption{$\alpha_1=0.3$}
  \end{subfigure}\\
  \begin{subfigure}[b]{0.24\linewidth}
    \begin{tikzpicture}
      \draw (0, 0) node[inner sep=0] {\includegraphics[width=\textwidth]{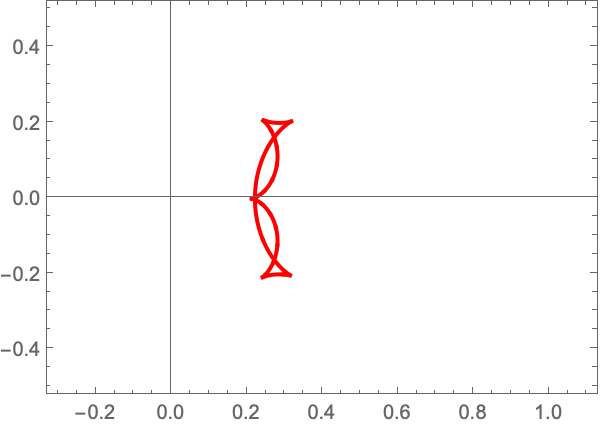}};
      \draw (0.6, 1.1) node {$3$};
      \draw (0.1, 0.7) node {$5$};
      \draw (0.1, 0.4) node {$1$};
      \draw (0.1, 0.1) node {$5$};
      \draw (0.1, -0.2) node {$1$};
      \draw (0.1, -0.5) node {$5$};
    \end{tikzpicture}
    \caption{$\alpha_1=0.32$}
  \end{subfigure}
  \begin{subfigure}[b]{0.24\linewidth}
    \begin{tikzpicture}
      \draw (0, 0) node[inner sep=0] {\includegraphics[width=\textwidth]{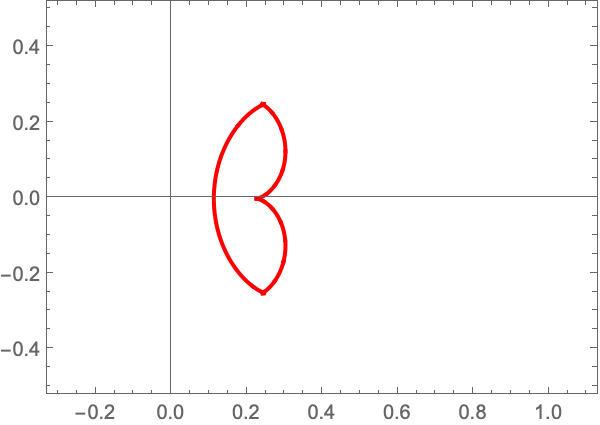}};
      \draw (0.6, 1) node {$3$};
      \draw (-0.45, 0.1) node {$1$};
    \end{tikzpicture}
    \caption{$\alpha_1=0.35$}
  \end{subfigure}
  \begin{subfigure}[b]{0.24\linewidth}
    \begin{tikzpicture}
      \draw (0, 0) node[inner sep=0] {\includegraphics[width=\textwidth]{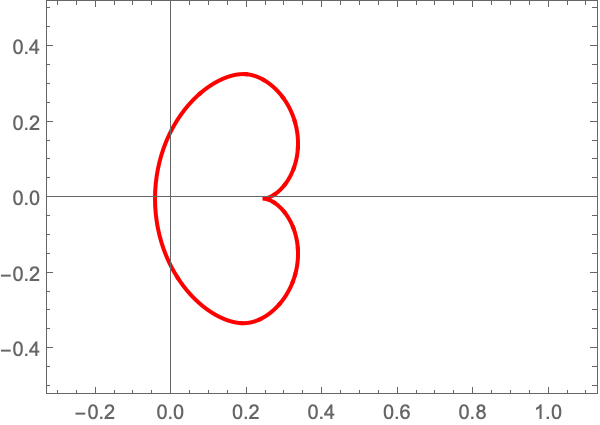}};
      \draw (0.6, 1) node {$3$};
      \draw (-0.45, 0.1) node {$1$};
    \end{tikzpicture}
    \caption{$\alpha_1=0.4$}
  \end{subfigure}
  \begin{subfigure}[b]{0.24\linewidth}
    \begin{tikzpicture}
      \draw (0, 0) node[inner sep=0] {\includegraphics[width=\textwidth]{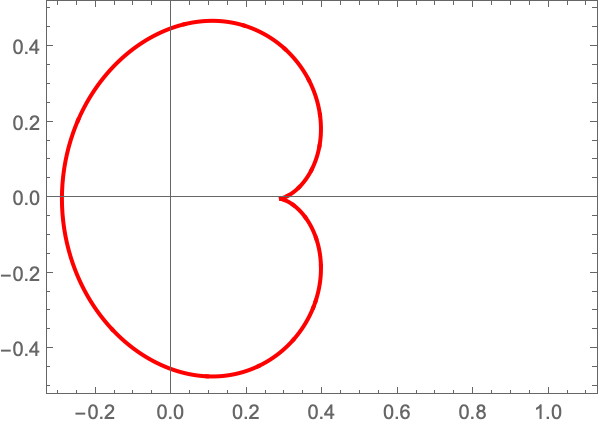}};
      \draw (0.6, 1) node {$3$};
      \draw (-0.45, 0.1) node {$1$};
    \end{tikzpicture}
    \caption{$\alpha_1=0.5$}
  \end{subfigure}
  \caption{The caustic curves for $\bm{\alpha}=(\alpha_1,0)$ for $\alpha_1=0,0.18,0.2, 0.3, 0.32,0.35,0.4,0.5$. The numbers indicate the number of images in each region bounded by the caustic curve.}\label{fig:caustics}
\end{figure*}

Including the frame-dragging effect breaks the radial symmetry of the lens and removes the degeneracy of the non-rotating point lens. The lens now has up to $5$ classical rays corresponding to stationary points of the time delay function. The critical and caustic curve gradually change as we increase the spin parameter $\alpha = \lVert \bm{\alpha}\rVert$ (see Fig.~\ref{fig:caustics}):
\begin{itemize}
  \item For $0 < \alpha < \frac{1}{3\sqrt{3}}\approx 0.192$, the critical curve consists of two loops. The corresponding caustics form a quadrangle and a triangle enclosing two $5$-image regions. Outside these caustic curves, there exist $3$ real classical rays.
  \item At $\alpha=\frac{1}{3 \sqrt{3}}$ the two $5$-image regions merge to form a larger $5$-image regions.
  \item At $\alpha \approx 0.31$, the right horizontal fold line starts to overtake the left horizontal caustics. When the right fold passes over the left fold, we observe the formation of two $1$-image regions.
  \item At $\alpha = \frac{1}{7} \sqrt{\frac{13+ 16 \sqrt{2}}{7}}\approx 0.322$ the two $1$-image regions merge.
  \item At $\alpha \approx 0.35$, the two remaining $5$-image regions collapse to a point in a swallowtail caustic. The $5$-image regions vanish for larger $\alpha$.
  \item For larger $\alpha$, we obtain a $1$-image region surrounded by a $3$-image region. The caustic curve consists of a fold curve and a single cusp point. As $\alpha$ increases, the caustic inflates. These findings for large $\alpha$ are in agreement with \cite{Sereno:2003}.
\end{itemize} 

For the rotating star, the outside region always consists of three classical rays. For each point $\bm{y}$ outside the caustics, the time delay function has three real and two complex critical points (forming a complex conjugate pair) in $\bm{x}$ when solving Eq.~\eqref{eq:variation}. When crossing a caustic while moving $\bm{y}$, two things can happen: the two complex critical points  merge on the real plane at the critical curve and become real classical rays forming a five-image region, or two real critical points merge on the critical curve and form a complex conjugate pair of complex saddle points. The complex critical points (sometimes known as complex rays) do not play a role in the geometric optics analysis of the rotating lens but do influence the interference pattern in wave optics (see for example \cite{Feldbrugge:2023}).

\subsection{Wave optics}
For a non-rotating point lens, the radial symmetry allows us to write the lensing amplitude as a radial integral, 
\begin{align}
  \Psi(\bm{y}) 
  &=\frac{w}{2\pi i} \int e^{i w \left(\frac{1}{2}(\bm{x}-\bm{y})^2 - \log x \right)}\mathrm{d}\bm{x}\\
  &=-ie^{i w y^2/2}\int_0^\infty J_0(w r y) e^{i w \left(\frac{1}{2}r^2  - \log r\right)}  r \mathrm{d}r\,,
\end{align}
using polar $\bm{x} = r (\cos \theta, \sin \theta)$ and Cartesian coordinates $\bm{y}=(y_1,y_2)$, the norm $y = \lVert \bm{y} \rVert$, and the integral representation of the Bessel function 
\begin{align}
  2\pi J_0\left( \sqrt{z_1^2+z_2^2}\right) = \int_0^{2\pi}e^{i (z_1 \cos \theta + z_2 \sin \theta)}\mathrm{d}\theta \,.
\end{align}
Remarkably, the radial integral can be evaluated
\begin{align}
  \Psi(\bm{y})  =2^{-1-\frac{i w }{2}} (-i w )^{1+\frac{i w }{2}} \Gamma \left(-\frac{i w}{2} \right) L_{-\frac{i w}{2}}\left(\frac{i w y^2 }{2}  \right)\label{eq:nonRotating}
\end{align}
in terms of the Laguerre function $L_n$ and the gamma function $\Gamma$. The intensity of the radiation assumes the form
\begin{align}
    |\Psi(\bm{y})|^2 = \frac{\pi w}{1-e^{-\pi w}}\left|_{1}F_{1}\left(\frac{i w}{2} , 1; \frac{i w y^2}{2} \right) \right|^2\,,
\end{align}
with the Kummer confluent hypergeometric function 
\begin{align}
    _{1}F_{1}(a,b;z) = \sum_{n=0}^\infty \frac{a^{(n)} z^n}{b^{(n)} n!}\,,
\end{align}
where we use the rising factorial $a^{(n)} = a (a+1) \dots (a+n-1)$ (see \cite{Nakamura:1999} for details).

\begin{figure*}
  \begin{subfigure}[b]{0.45\linewidth}
    \includegraphics[width=\textwidth,height=0.65\textwidth]{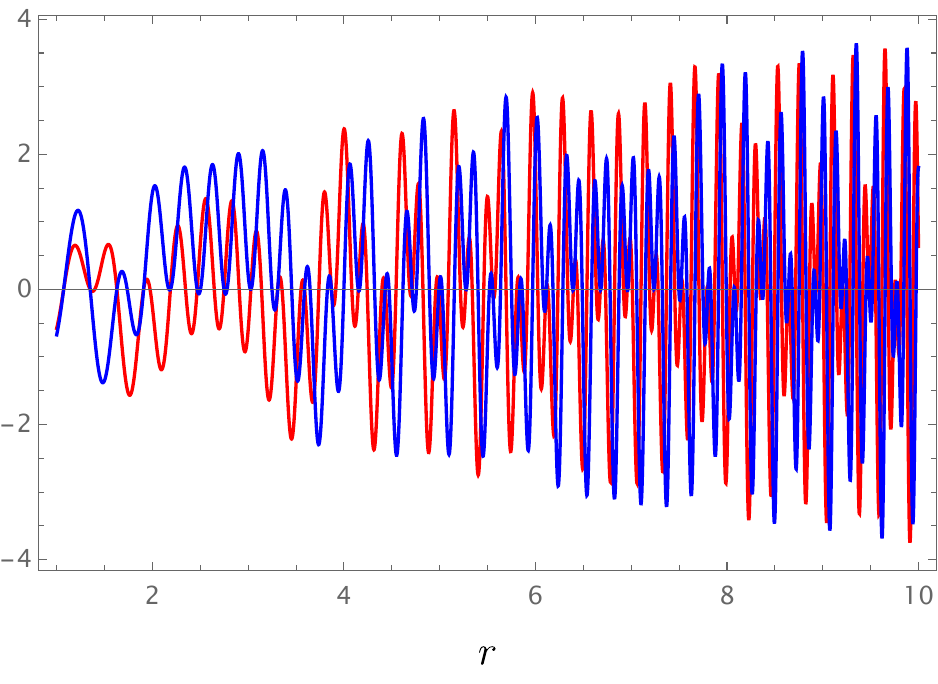}
    \caption{Original contour}
  \end{subfigure}
  \begin{subfigure}[b]{0.45\linewidth}
    \includegraphics[width=\textwidth,height=0.64\textwidth]{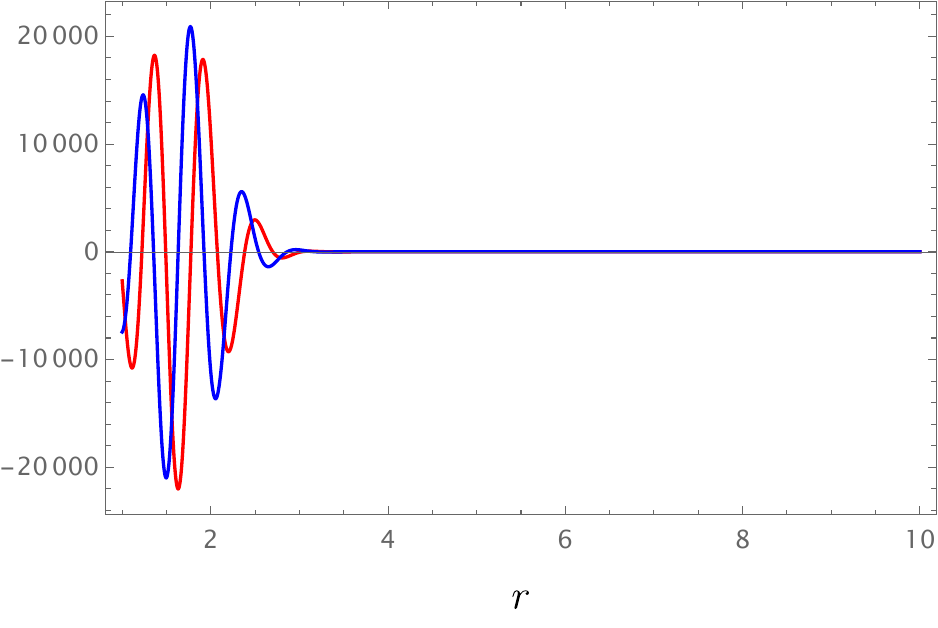}
    \caption{Deformed contour}
  \end{subfigure}
  \caption{The integrand of the radial integral in Eq.~\eqref{eq:KF_rot} along the real line $[1,\infty)$ (left), and deformed contour $[1, e^{i \pi/4} \infty)$. The blue and red lines indicate the real and imaginary parts, respectively.}\label{fig:deformation}
\end{figure*}

For a rotating lens, the frame-dragging term in polar coordinates is
\begin{align}
  \frac{\bm{\alpha} \cdot \bm{x}}{x^2} = \frac{\alpha_1  \cos \theta + \alpha_2 \sin \theta}{r}\,,
\end{align}
with $\bm{\alpha}=(\alpha_1,\alpha_2)$. This makes the lensing amplitude for the non-rotating lens take on the following form
\begin{align}
  \Psi(\bm{y})  &= -i w e^{i w y^2/2}\nonumber\\
  &\times  \int_{0}^{\infty} 
  J_0\left(w r \lVert\bm{y} - \bm{\alpha}/r^2\rVert \right) 
  e^{i w(r^2/2-\log r)}r \mathrm{d}r\,.\label{eq:KF_rot}
\end{align}
Note that this is only a slight variation on the amplitude evaluated by \citep{Baraldo:1999} using the approximation in Eq.~\eqref{eq:approx}:
\begin{align}
  \Psi(\bm{y}) \sim&  -i w e^{i w y^2/2}\nonumber\\
  &  \times  \int_{0}^{\infty} 
  J_0\left(w r \lVert\bm{y} - \bm{\alpha}\rVert \right) 
  e^{i w(r^2/2-\log r)}r \mathrm{d}r \, ,
\end{align}
which can be evaluated analytically as this is simply the Kirchoff-Fresnel integral for the non-rotating lens but shifted
\begin{equation}
     \Psi (\bm{y}) \stackrel{\text{\cite{Baraldo:1999}}}{=}2^{-1-\frac{i w }{2}} (-i w )^{1+\frac{i w }{2}} \Gamma \left(-\frac{i w}{2} \right) L_{-\frac{i w}{2}}\left(\frac{i w \lVert \bm{y}- \bm{\alpha}\rVert^2 }{2}  \right)\,.
\end{equation}
Unfortunately, the radial integral in Eq.~\eqref{eq:KF_rot} cannot be evaluated using special functions and is highly oscillatory for large $r$. However, as the integrand is analytic, and the analytic continuation is dominated by the Gaussian term $e^{iw r^2/2}$ for large $|r|$ in the complex plane, we can safely deform the half line $[0,\infty)$ to an integration contour starting at the origin $0$ tangential to the real axis and ending at $e^{i\pi/4} \infty$. This deformation suppresses the integrand for large $|r|$ and enables the efficient evaluation of the amplitude $\Psi(\bm{y})$ using conventional integration techniques. As the analytic continuation diverges around $r=0$ in the complex plane, we deform the real half-line $(0,\infty)$ into the two line segments $(0,1] \cup [1,e^{i \pi /4} \infty)$ (see Fig.~\ \ref{fig:deformation}). This is a rudimentary application of Picard-Lefschetz theory \cite{Feldbrugge:2023}. More intricate deformations of the original integration domain can further improve the convergence of the radial integral. However, this simple deformation suffices for the purpose of this paper.

\subsection{Validating the lens amplitude}
To validate our numerical method for the evaluation of the wave amplitude in the previous section, we approximate the Kirchhoff-Fresnel integral in three ways. 

\begin{figure*}
  \centering
  \begin{subfigure}[b]{0.32\linewidth}
    \includegraphics[width=\textwidth]{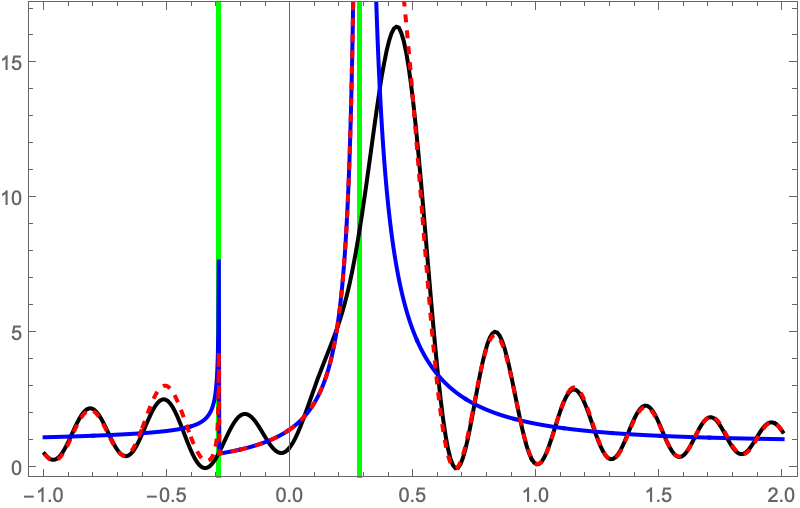}
    \caption{$w=10$}
  \end{subfigure}
  \begin{subfigure}[b]{0.32\linewidth}
    \includegraphics[width=\textwidth]{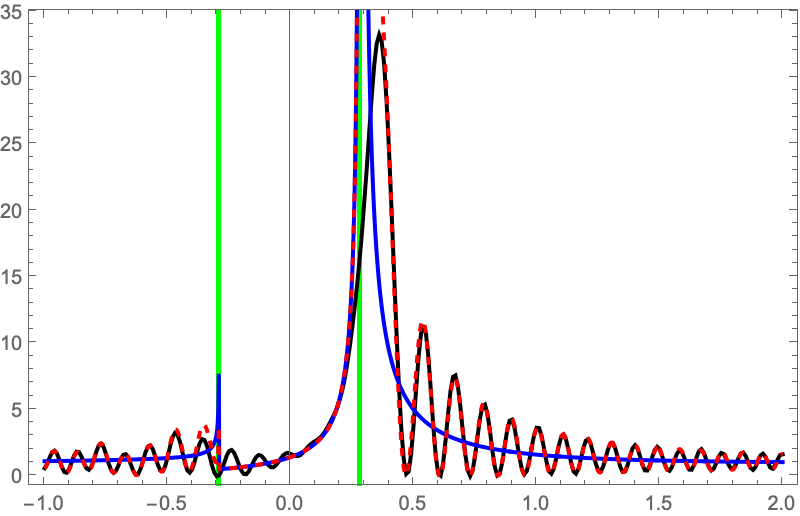}
    \caption{$w=30$}
  \end{subfigure}
  \begin{subfigure}[b]{0.32\linewidth}
    \includegraphics[width=\textwidth]{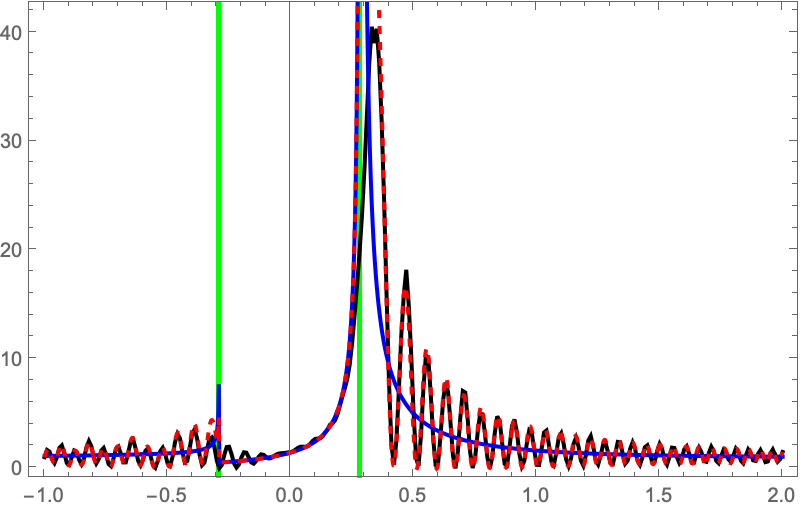}
    \caption{$w=50$}
  \end{subfigure}
  \caption{The intensity $|\Psi(\bm{y})|^2$ for $y$ along the equatorial plane/for $y=(y_1,0)$  with $-1 \leq y_1 \leq 2$ for $\bm{\alpha}=(1/2,0)$ in geometric optics (blue), the Eikonal approximation (red), and the numerical evaluation of the Kirchhoff-Fresnel integral (black). The vertical green lines mark the caustics separating the one-image region (between the two green lines) from the triple-image regions (the regions on the outsides of the green lines).}\label{fig:Eikonal}
\end{figure*}

First, the Eikonal approximation of $\Psi(\bm{y})$ is a good approximation in the semi-classical regime, bridging the geometric optics approximation \eqref{eq:geometric} and the full Kirchhoff-Fresnel integral,
\begin{align}
    \Psi(\bm{y}) \approx \sum_{\bm{x} \in \bm{\xi}^{-1}(\bm{y})} \frac{e^{i \omega T(\bm{x},\bm{y}) - i n(\bm{x}) \pi /2}}{\sqrt{|\det \nabla \bm{\xi}(\bm{x})|}}\,,
\end{align}
with the Morse index $n$ of the critical point $\bm{x}$ ($0$ for minima, $1$ for saddle points, and $2$ for maxima of the time delay $T$). For a detailed exposition of the Eikonal approximation see \cite{Schneider:1992}. In the present discussion, we will only include the real rays, though the Eikonal approximation can be extended by including relevant complex rays \cite{Feldbrugge:2023}. In Fig.~\ \ref{fig:Eikonal}, we compare the geometric optics approximation, the Eikonal approximation and the Kirchhoff-Fresnel integral for a rotating lens. The geometric optics approximation captures the main  behavior of the lensing pattern but misses the interference of the three real rays in the triple-image regions. Additionally, the geometric optics approximation diverges at the caustics. The Eikonal approximation is a significant improvement as it captures the interference in the triple-image regions. However, it fails to capture the oscillations on the left side of the single-image region. These oscillations result from the interference of the real ray and a complex ray (associated with the left fold caustic). Both the geometric optics and Eikonal approximations overestimate the intensity near the left cusp caustic. Given that the oscillations of the Eikonal approximation and the numerical evaluation of the Kirchhoff-Fresnel integral line up, and the deviations can be understood, we are confident about the accuracy of the numerical evaluation.

\begin{figure*}
  \centering
  \begin{subfigure}[b]{0.49\linewidth}
    \includegraphics[width=\linewidth]{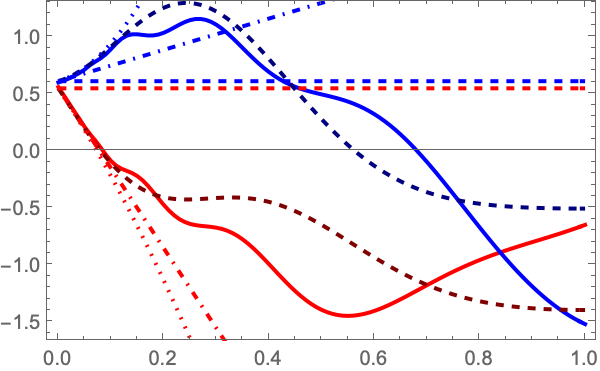}
    \caption{$w=10$}
  \end{subfigure}
  \begin{subfigure}[b]{0.49\linewidth}
    \includegraphics[width=\linewidth]{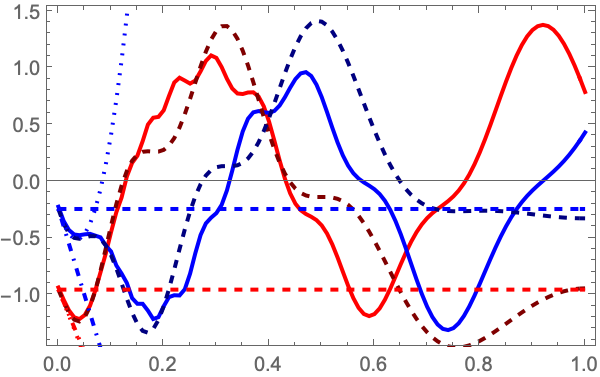}
    \caption{$w=30$}
  \end{subfigure}
  \caption{A comparison of the exact evaluation (solid curves) and approximations of the lens amplitude as a function of $\bm{\alpha}=(\alpha,0)$ for $\omega=10$ (left panel) and $\omega = 30$ (right panel) and $\bm{y}=(1,1)$. We compare the numerical evaluation with the zeroth-order (dashed curves), first-order (dashed-dotted curves), second-order approximation (dotted curves), and the shifted non-rotating star (dark dashed curves). The real and imaginary parts of $\Psi$ are plotted in red and blue, respectively.}\label{fig:Perturbations}
\end{figure*}

The second method to validate our numerical results is by expanding the radial integral in the Kirchhoff-Fresnel integral \eqref{eq:KF_rot} in powers of the rotation parameter, for which we can obtain analytic results. As the expansion of the Bessel function, 
\begin{widetext}
\begin{align}
    &J_0\left(w r  \lVert \bm{y} -\bm{\alpha}/r^2\rVert\right) \\
    &= J_0(w r y) 
    + w \bm{\alpha}\cdot \bm{y}  \frac{J_1(w r y)}{r y}
    - w^2 (\bm{\alpha} \cdot \bm{y})^2 \frac{J_0(w r y)}{2 r^2 y^2}
    + w \left(y_1^2 (\alpha_1^2-\alpha_2^2)+4 \alpha_1 \alpha_2 y_1 y_2-y_2^2 (\alpha_1^2-\alpha_2^2) \right)\frac{J_1(w r y)}{2 r^3 y^3}
    +\mathcal{O}(\bm{\alpha}^3)\,,\nonumber
\end{align}
only features the Bessel functions $J_0(w r y)/ r^m$ and $J_1(w r y)/ r^m$ for positive integers $m$, the closed form integral 
\begin{align}
  &\int_0^\infty J_\nu(w y  r) e^{i w \left(\frac{1}{2}r^2 - \log r\right)}r^{n}\mathrm{d}r\nonumber\\
  &=2^{-\frac{1+\nu-n+i w }{2} } (-i w )^{-\frac{1+\nu+n-i w }{2} } \left(w y  \right)^{\nu} \Gamma \left(\frac{1+ \nu+n-i w }{2}\right)
  \, _1\tilde{F}_1\left(\frac{1 + \nu+n-i w}{2};1 + \nu;-\frac{i y^2 w}{2} \right)    
\end{align}
allows us to evaluate the radial integral and approximate the integral to any order in $\bm{\alpha}$. Fig.~\ref{fig:Perturbations} shows a comparison of this perturbative expansion and the numerically evaluated Kirchhoff-Fresnel integral. The expansion matches the numerical evaluation well for small $\lVert \bm{\alpha} \rVert$ but requires many terms to capture the oscillations of the interference pattern.
\end{widetext}

Finally, the numerical evaluation matches qualitatively the shifted interference pattern of the non-rotating star for small $\alpha$ (see the next section for details).

\section{Results}\label{sec:results}
\begin{figure*}
  \begin{subfigure}[b]{0.32\linewidth}
      \includegraphics[width=\textwidth]{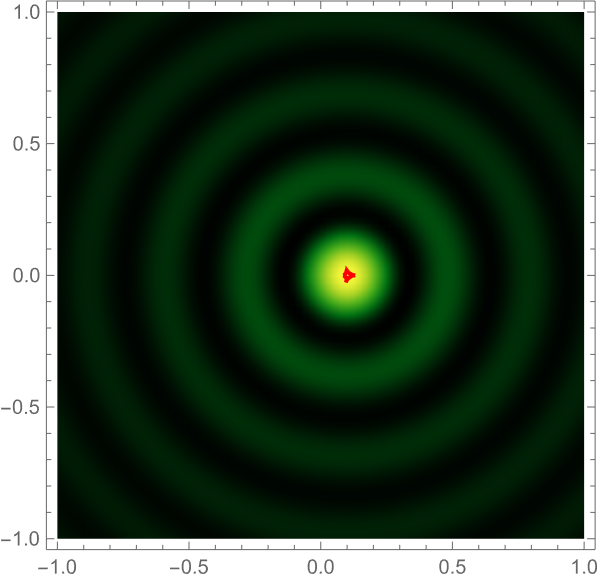}
  \end{subfigure}
  \begin{subfigure}[b]{0.32\linewidth}
      \includegraphics[width=\textwidth]{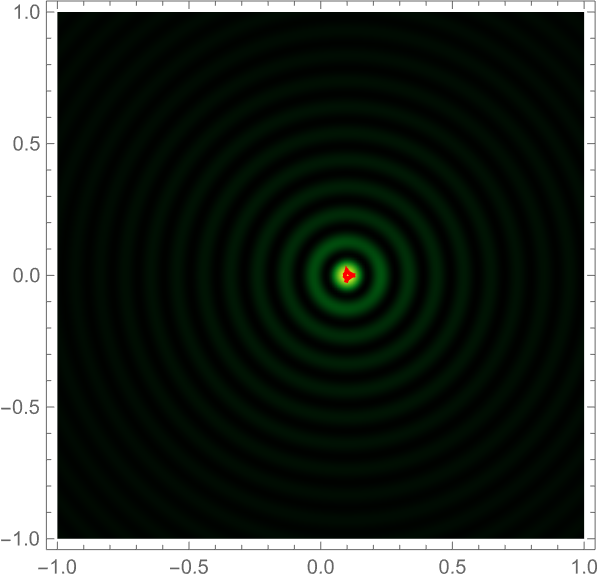}
  \end{subfigure}
  \begin{subfigure}[b]{0.32\linewidth}
      \includegraphics[width=\textwidth]{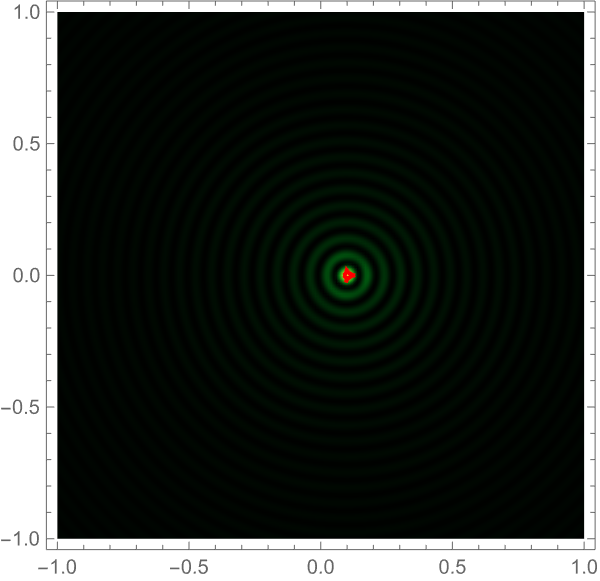}
  \end{subfigure}\\
  \begin{subfigure}[b]{0.32\linewidth}
      \includegraphics[width=\textwidth]{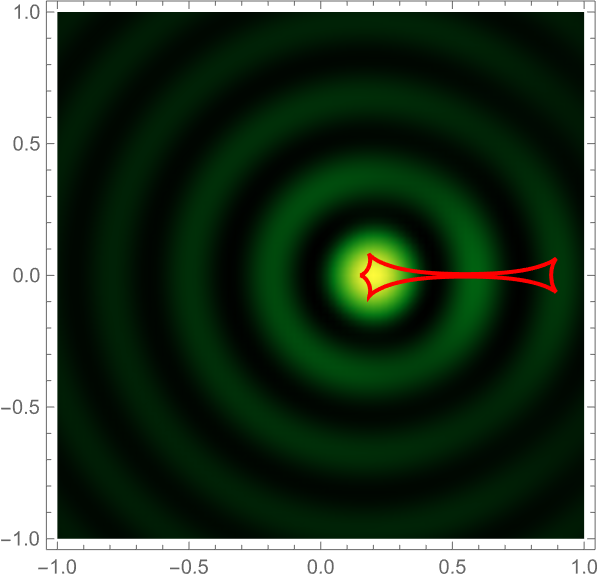}
  \end{subfigure}
  \begin{subfigure}[b]{0.32\linewidth}
      \includegraphics[width=\textwidth]{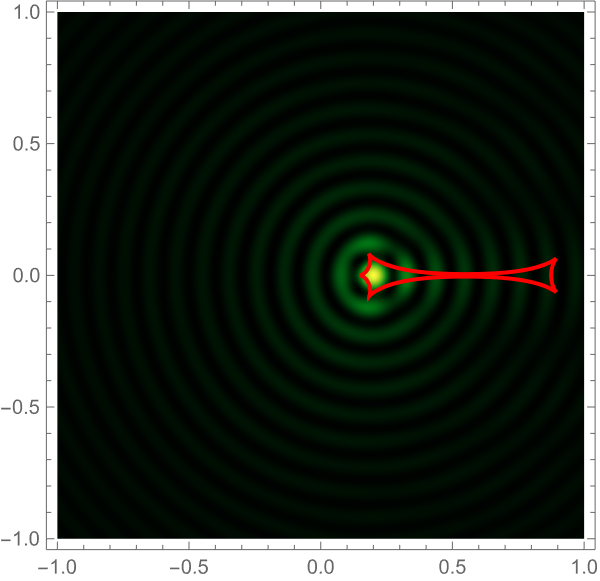}
  \end{subfigure}
  \begin{subfigure}[b]{0.32\linewidth}
      \includegraphics[width=\textwidth]{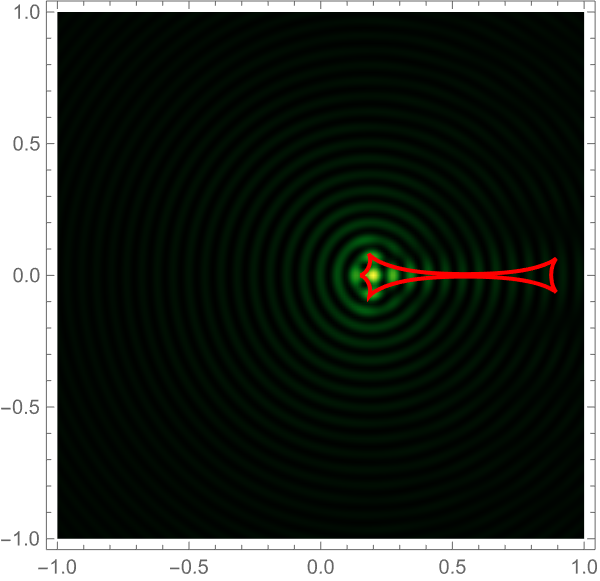}
  \end{subfigure}\\
  \begin{subfigure}[b]{0.32\linewidth}
      \includegraphics[width=\textwidth]{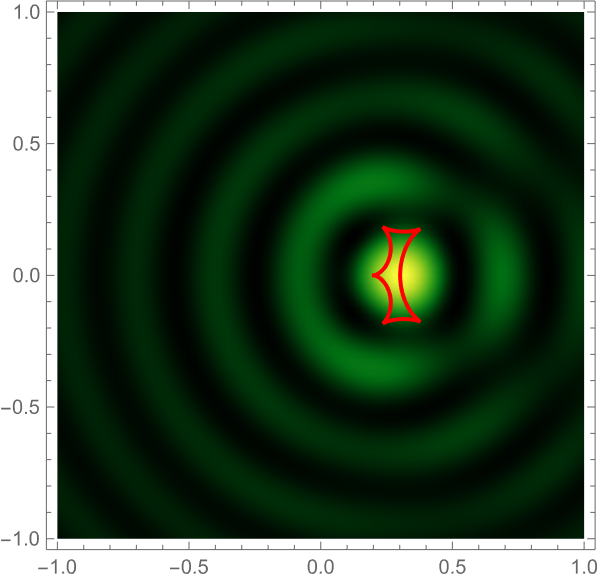}
  \end{subfigure}
  \begin{subfigure}[b]{0.32\linewidth}
      \includegraphics[width=\textwidth]{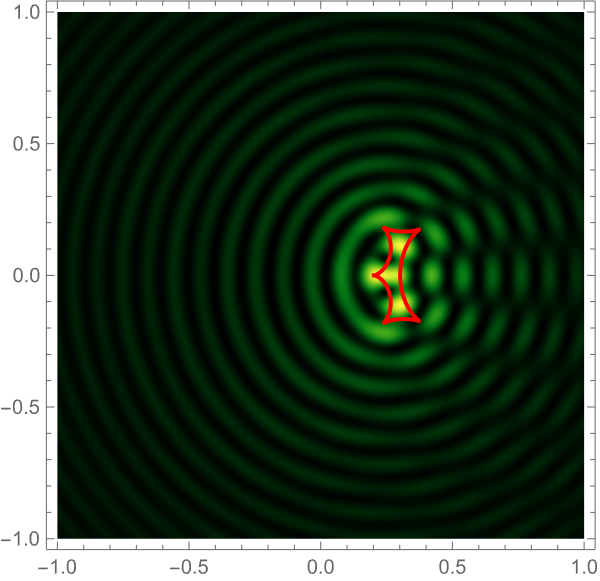}
  \end{subfigure}
  \begin{subfigure}[b]{0.32\linewidth}
      \includegraphics[width=\textwidth]{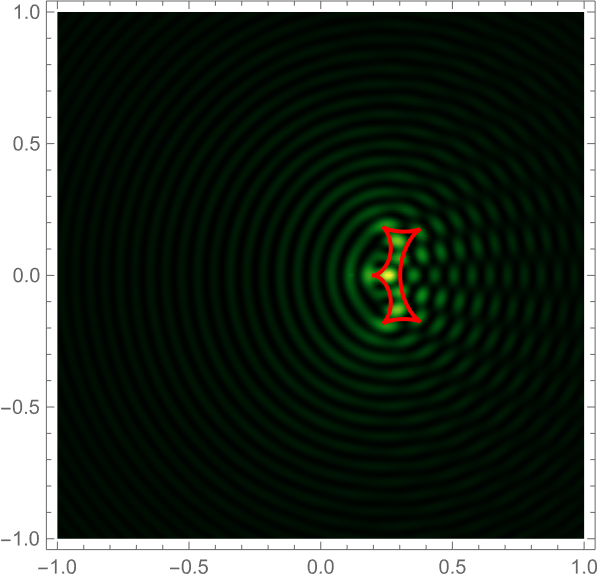}
  \end{subfigure}\\
  \begin{subfigure}[b]{0.32\linewidth}
      \includegraphics[width=\textwidth]{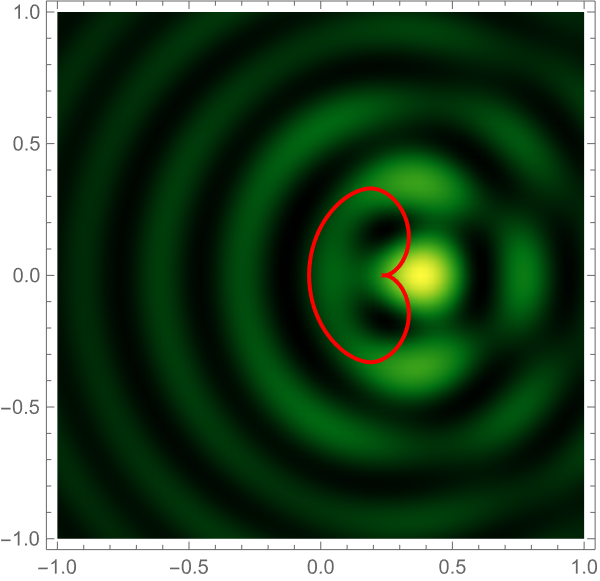}
  \end{subfigure}
  \begin{subfigure}[b]{0.32\linewidth}
      \includegraphics[width=\textwidth]{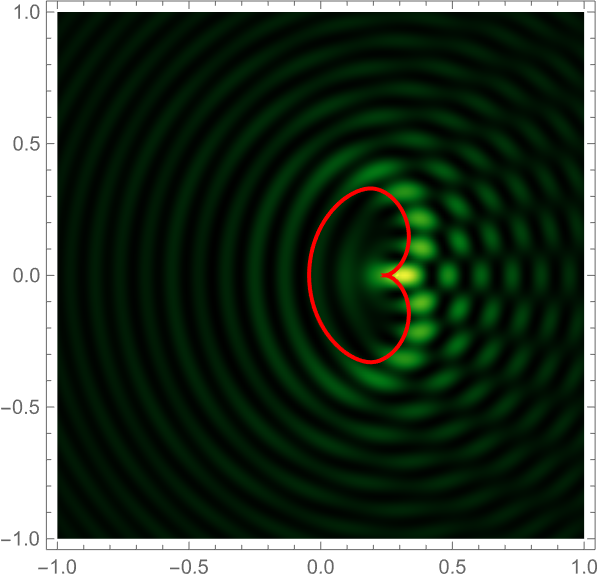}
  \end{subfigure}
  \begin{subfigure}[b]{0.32\linewidth}
      \includegraphics[width=\textwidth]{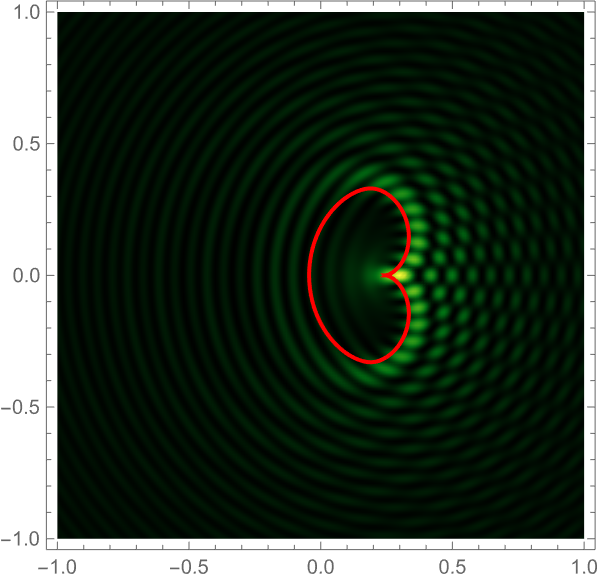}
  \end{subfigure}
  \caption{Interference patterns of the rotating lens for $\bm{\alpha}=(\alpha_1,0)$ for $\alpha_1=0.1, 0.2, 0.3, 0.4$ (top to bottom) and the angular frequencies $w=10,30,50$ (left to right). The red curves are the caustics: in the geometric optics approximation, the intensity becomes infinite there.}\label{fig:interference}
\end{figure*}

When evaluating the Kirchhoff-Fresnel pattern as a function of $\bm{y}$ we obtain an interference pattern matching the caustics obtained from the geometric optics approximation (see Fig.~\ \ref{fig:interference}). For small $\lVert \bm{\alpha}\rVert$, the caustic curve is small compared to the typical length scale of the interference pattern and the interference pattern is close to the shifted pattern predicted by \cite{Baraldo:1999} (although see below for an explicit comparison). For larger $\lVert \bm{\alpha}\rVert$, the interference pattern becomes more intricate neatly following the caustics. Note that the oscillations in the interference pattern in the single image region for $\lVert \bm{\alpha} \rVert \geq 0.322$ are the result of a relevant complex ray corresponding to a complex critical point of the time delay function.

\begin{figure*}
  \centering
  \begin{subfigure}[b]{0.32\linewidth}
      \includegraphics[width=\textwidth]{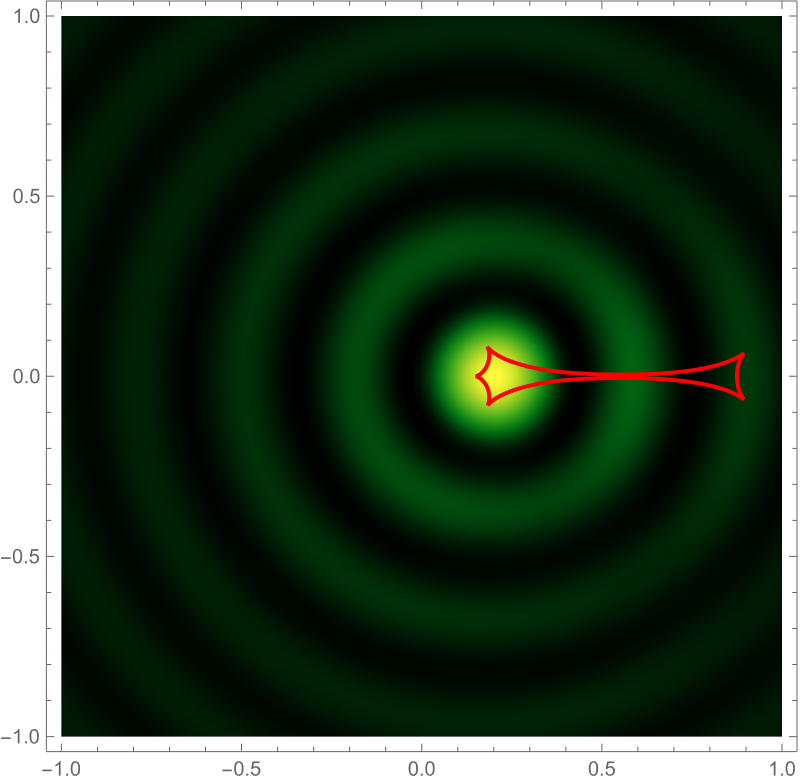}
  \end{subfigure}
  \begin{subfigure}[b]{0.32\linewidth}
      \includegraphics[width=\textwidth]{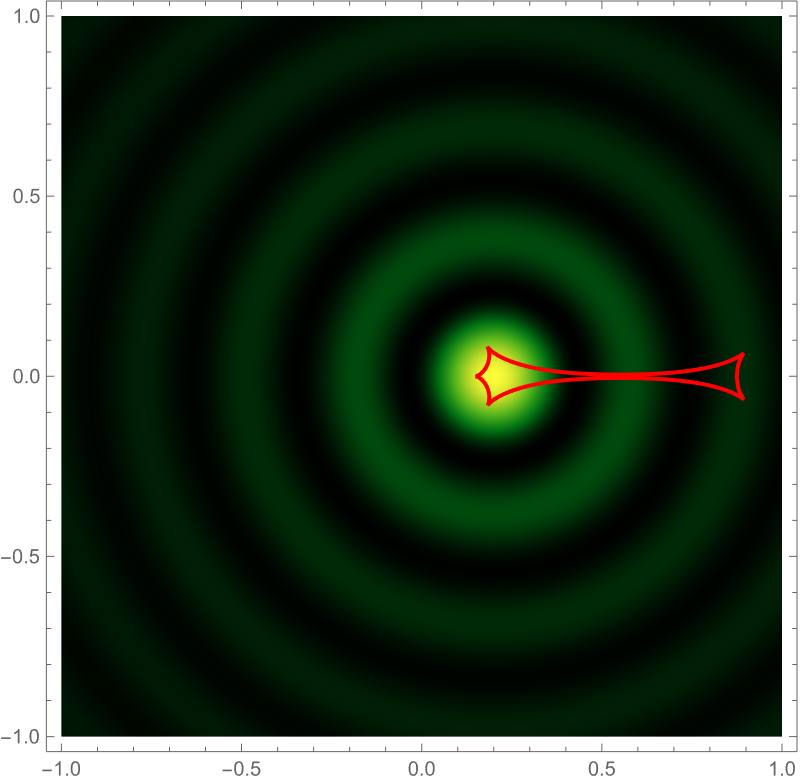}
  \end{subfigure}
  \begin{subfigure}[b]{0.32\linewidth}
      \includegraphics[width=\textwidth]{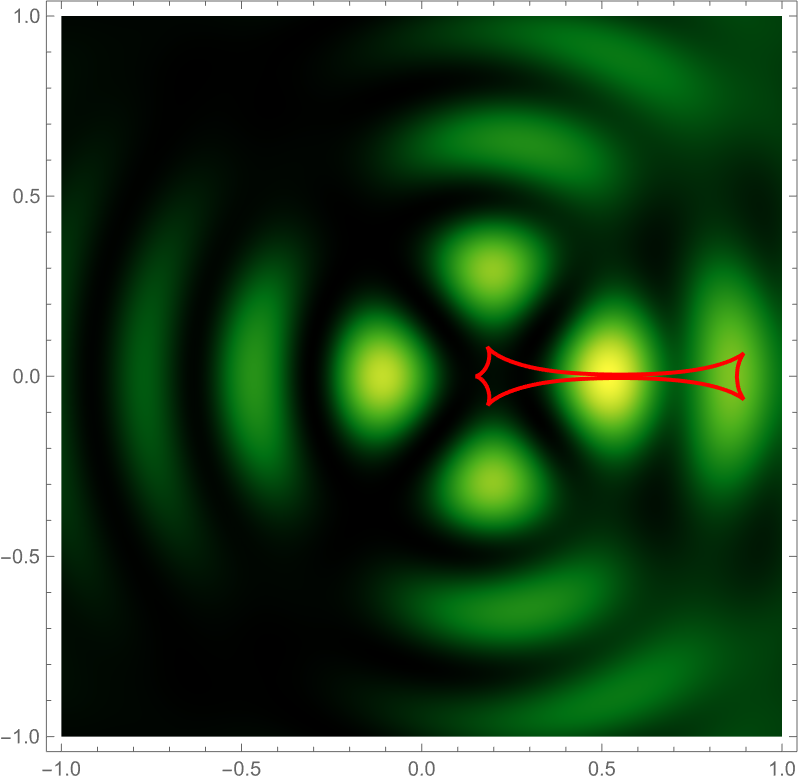}
  \end{subfigure}\\
  \begin{subfigure}[b]{0.32\linewidth}
      \includegraphics[width=\textwidth]{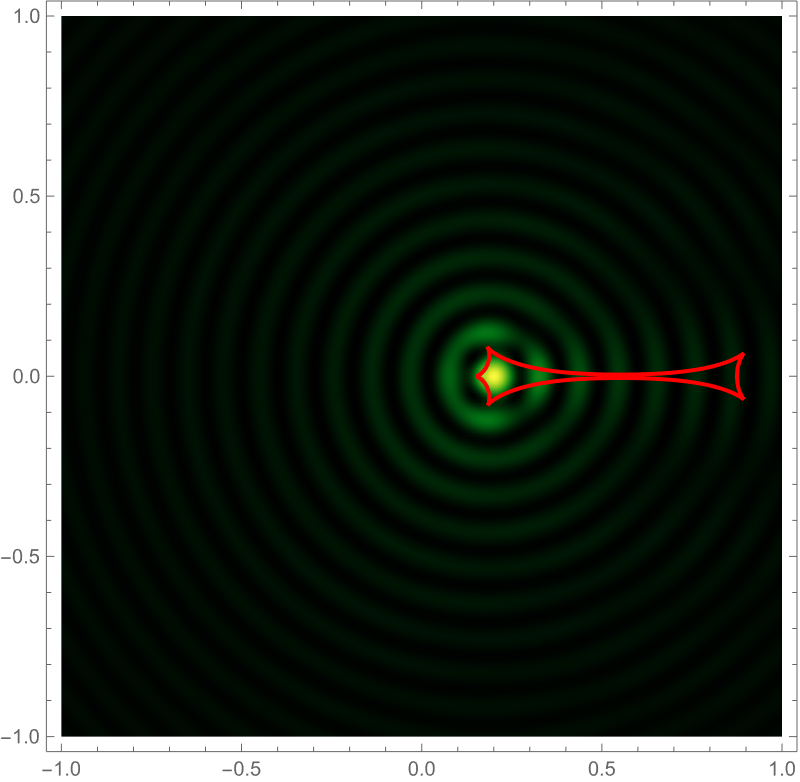}
  \end{subfigure}
  \begin{subfigure}[b]{0.32\linewidth}
      \includegraphics[width=\textwidth]{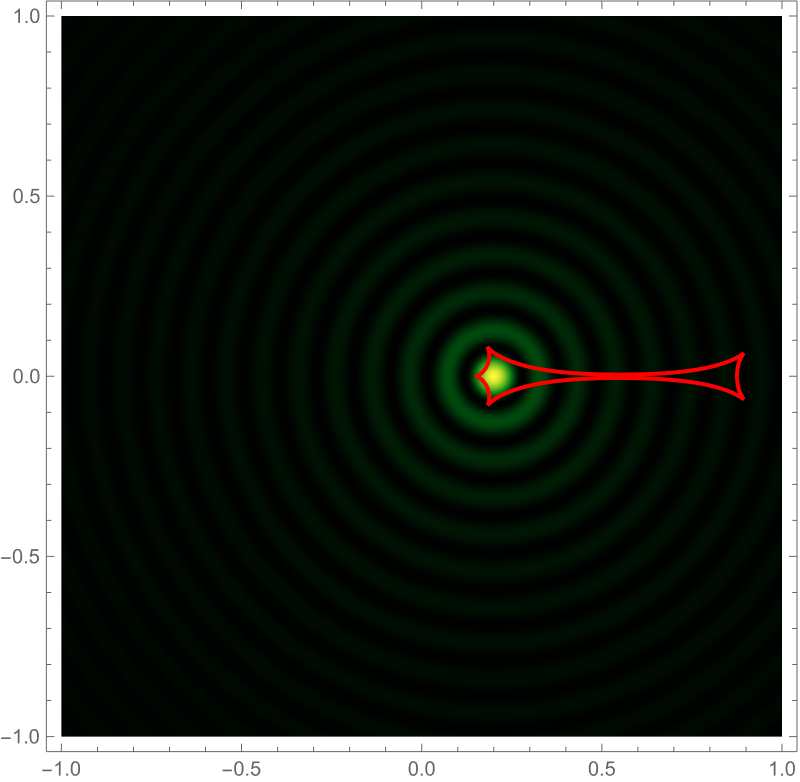}
  \end{subfigure}
  \begin{subfigure}[b]{0.32\linewidth}
      \includegraphics[width=\textwidth]{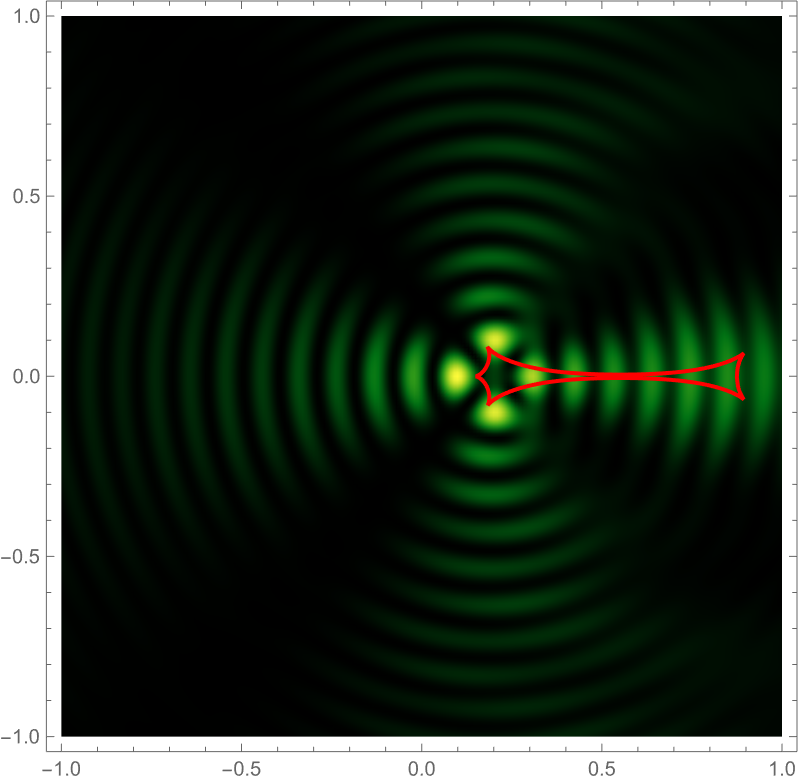}
  \end{subfigure}
  \caption{A comparison of the exact Kirchhoff-Fresnel integral (left) and the shifted non-rotating interference pattern (center) and their difference (right) for $\bm{\alpha}=(0.2,0)$, for $w=10$ (the upper panels) and $w=30$ (the lower panels). The difference between the two patterns is reminiscent of the light emitted by a lighthouse.} \label{fig:comparison}
\end{figure*}

We compare the interference pattern with the shifted non-rotating lens proposed by \cite{Baraldo:1999} in Fig.~\ \ref{fig:comparison}. Though the two interference patterns appear similar, they show a systematic difference even for small $\alpha$: this difference is highlighted in the right column. This difference can in principle allow one to directly infer the spin of the rotating start through frame dragging from the observed fringes. 

We want to highlight that using the shifted non-rotating time delay to compute the interference pattern generated by the rotating star is \emph{incorrect} (see also the discussion around Eq.~\eqref{eq:expansion} in Sec.~\eqref{sec:time-delay}). The underlying reason why the approximation 
\begin{subequations}\label{eq:time-delay-approximation}
\begin{align}\label{eq:time-delay-approximation:1}
  T(\bm{x},\bm{y}) 
 & =\frac{(\bm{x}-\bm{y})^2}{2}  -\log \lVert \bm{x} - \bm{\alpha}\rVert - \sum_{n=2}^\infty \frac{\alpha^n \cos n \bar{\theta}}{n x^n} \\\label{eq:time-delay-approximation:2}
 &\approx  \frac{(\bm{x}-\bm{y})^2}{2}  -\log \lVert \bm{x} - \bm{\alpha}\rVert 
  \,,  
\end{align}\end{subequations} 
fails despite its seeming correctness (and its extensive usage in the literature~\cite{Baraldo:1999,Asada:2000vn,Sereno:2002tv,Ebrahimnejad:2005}) is that the residue $\sum_{n=2}^\infty \frac{\alpha^n \cos n \bar{\theta}}{n x^n}$ diverges for $\bm{x}=\bm{0}$, and $\bm{x}=\bm{\alpha}$ (following from the fact that we approximate a function $\log \lVert \bm{x}\rVert$ which diverges at $\bm{x}=\bm{0}$ by the function $\log\lVert \bm{x} - \bm{\alpha}\rVert$ which diverges at $\bm{x}=\bm{\alpha}$). The time delays \eqref{eq:time-delay-approximation:1} and \eqref{eq:time-delay-approximation:2} give rise to radically different interference patterns. This can be seen by estimating the interference pattern generated by the correction term $\sum_{n=2}^\infty \frac{\alpha^n \cos n \bar{\theta}}{n x^n}$. The Kirchhoff-Fresnel integral for the time delay~\eqref{eq:time-delay-approximation:1} is
\begin{equation}\label{eq:KF-full}
     \Psi(\bm{y}) 
      = \frac{w}{2\pi i} \int e^{i w \left[\frac{(\bm{x}-\bm{y})^2}{2}  -\log \lVert \bm{x} - \bm{\alpha}\rVert
      - \sum_{n=2}^\infty \frac{\alpha^n \cos n \bar{\theta}}{n x^n}\right]}\mathrm{d}\bm{x}\,.
\end{equation} Using the identity $e^x = \sum_{k=0}^\infty \frac{x^k}{k!}$, we obtain
\begin{equation}\label{eq:delta-psi}
\begin{split}
     & \Delta  \Psi(\bm{y})  = \Psi(\bm{y})- \Psi_{\text{shifted}}(\bm{y}) \\
    &=- \frac{w^2}{2\pi } \int e^{i w \left[\frac{(\bm{x}-\bm{y})^2}{2}  -\log \lVert \bm{x} - \bm{\alpha}\rVert\right]} \frac{\alpha^2 \cos 2\bar{\theta}}{2x^2}\mathrm{d}\bm{x} +\mathrm{O}(\alpha^3) 
\end{split}
\end{equation} where $\Psi(\bm{y})$ is given by Eq.~\eqref{eq:KF-full} and $\Psi_{\text{shifted}}$ is 
\begin{equation}
    \Psi_{\text{shifted}}(\bm{y}) 
      = \frac{w}{2\pi i} \int e^{i w \left[\frac{(\bm{x}-\bm{y})^2}{2}  -\log \lVert \bm{x} - \bm{\alpha}\rVert
      \right]}\mathrm{d}\bm{x}\,.
\end{equation} For the purpose of the argument, it is enough to consider the correction of order $\alpha^2$ to the shifted non-rotating interference pattern. If considering the approximation to the time delay given by Eq.~\eqref{eq:time-delay-approximation:2}  was correct, then the next order correction in $\alpha$ to the time delay in Eq.~\eqref{eq:delta-psi} either 1) does not significantly impact the interference pattern, or 2) if it does, the intensity of the correction is smaller than roughly $\alpha^2$. As we will see in the following, neither of these two statements is true when evaluating Eq.~\eqref{eq:delta-psi}.  Recall that $\bar{\theta}$ is a function of $\bm{x}$ and $\bm{\alpha}$, in particular
\begin{equation}
    \frac{\alpha^2\cos 2\bar{\theta}}{2x^2} = \frac{2(\bm{x}\cdot \bm{\alpha})-\alpha^2 x^2}{2x^4}\,.
\end{equation} 
Using polar coordinates centered at $\bm{x}=\bm{\alpha}$\begin{equation}\label{eq:transformation-polar-coordinates}
    \bm{x}-\bm{\alpha} = r(\cos\theta\,,\sin\theta)\,,
\end{equation} 
we can rewrite expression~\eqref{eq:delta-psi} as
\begin{equation}
\begin{split}
   & \Delta\Psi =- \frac{w^2 e^{iw \frac{(\bm{\alpha}-\bm{y})^2}{2}}}{4\pi } \int \mathrm{d}\theta \left[(\alpha_1^2-\alpha_2^2) \cos2\theta +2\alpha_1\alpha_2 \sin 2\theta\right]\times\\
   & \int \mathrm{d} r  e^{i w \left[\frac{r^2}{2}+\frac{(\bm{\alpha}-\bm{y})^2}{2} +r(\alpha_1-y_1)\cos\theta+r(\alpha_2-y_2)\cos\theta\right]}  r^{-1-iw}+\mathrm{O}(\alpha^3) 
\end{split}    
\end{equation} where we only kept terms up to second order in $\alpha^2$, \textit{i.e.},
\begin{equation}
     \frac{\alpha^2\cos 2\bar{\theta}}{2x^2} = \frac{(\alpha_1^2-\alpha_2^2)\cos 2\theta +2\alpha_1\alpha_2 \sin 2\theta}{2r^2} +\mathrm{O}(\alpha^3)\,.
\end{equation} In practice, this means that for this term we can use the transformation $\bm{x} =r(\cos\theta,\sin\theta)$ rather than Eq.~\eqref{eq:transformation-polar-coordinates} since the factor of $\bm{\alpha}$ in the transformation enters as a higher order correction.

The radial integral 
\begin{equation}
    I_1=\int_0^\infty \mathrm{d} r  e^{i w \left[\frac{r^2}{2}+\frac{(\bm{\alpha}-\bm{y})^2}{2} +r b\right]}  r^{-1-iw}\,,
\end{equation} 
regularized using analyticity \cite{Feldbrugge:2023c}, can be evaluated in terms of the Kummer's confluent hypergeometric function
\begin{widetext}
    \begin{equation}
    I_1 = 2^{-1-iw/2} e^{\pi w/4} w^{i w/2} \left\{(i-1) b\sqrt{w} \Gamma\left[\frac{1}{2}-\frac{iw}{2}\right] {}_1F_1\left[\frac{1}{2}-\frac{iw}{2}\,, \frac{3}{2}\,, -\frac{iw}{2} b^2 \right] +\Gamma\left[-\frac{iw}{2}\right] {}_1 F_1 \left[-\frac{iw}{2}\,,\frac{1}{2}\,,-\frac{iw}{2} b^2\right]\right\}\,.
\end{equation}
\begin{figure}
    \centering
    \includegraphics[width=0.5\linewidth]{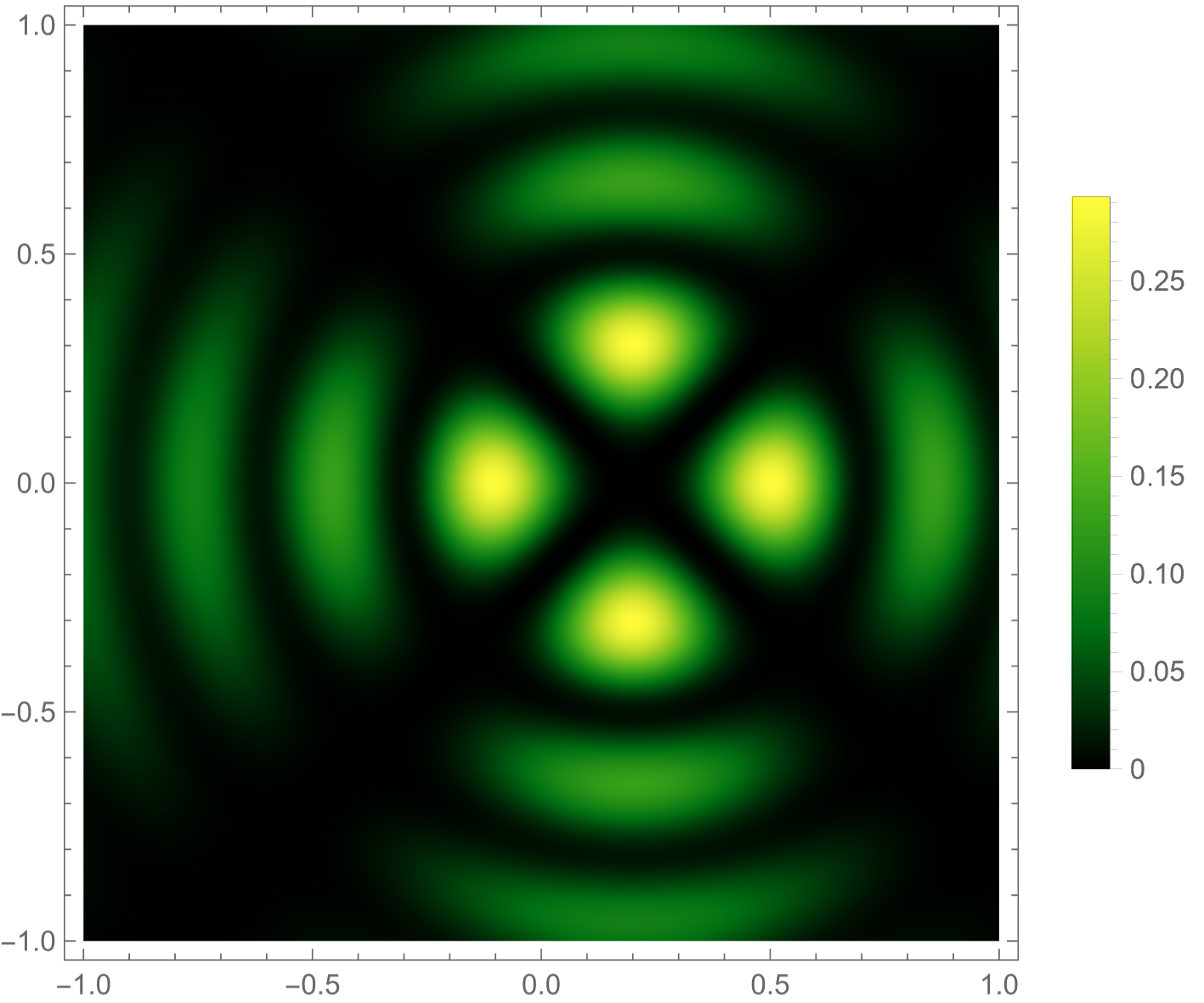}~\includegraphics[width=0.5\linewidth]{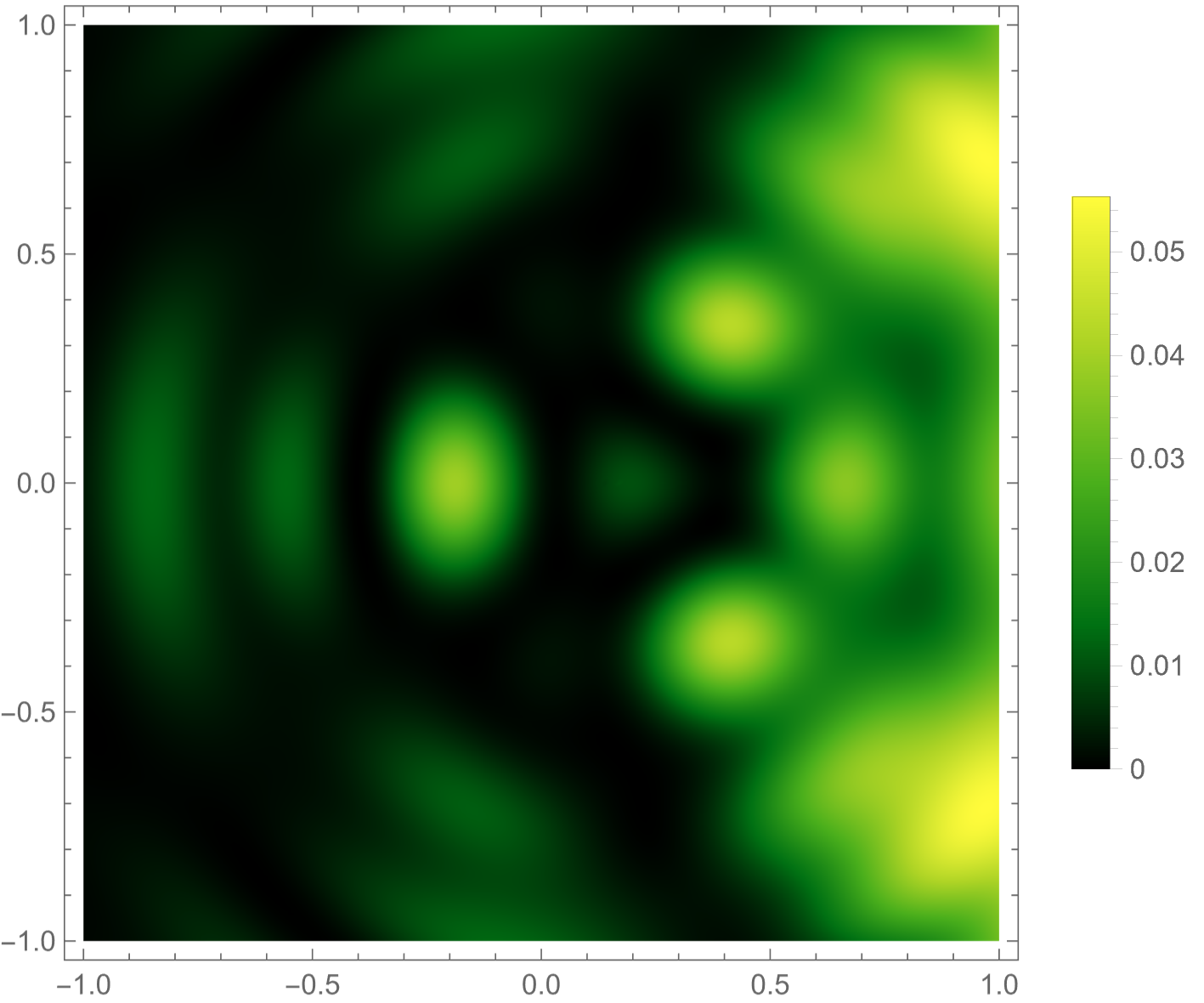}
    \includegraphics[width=0.5\linewidth]{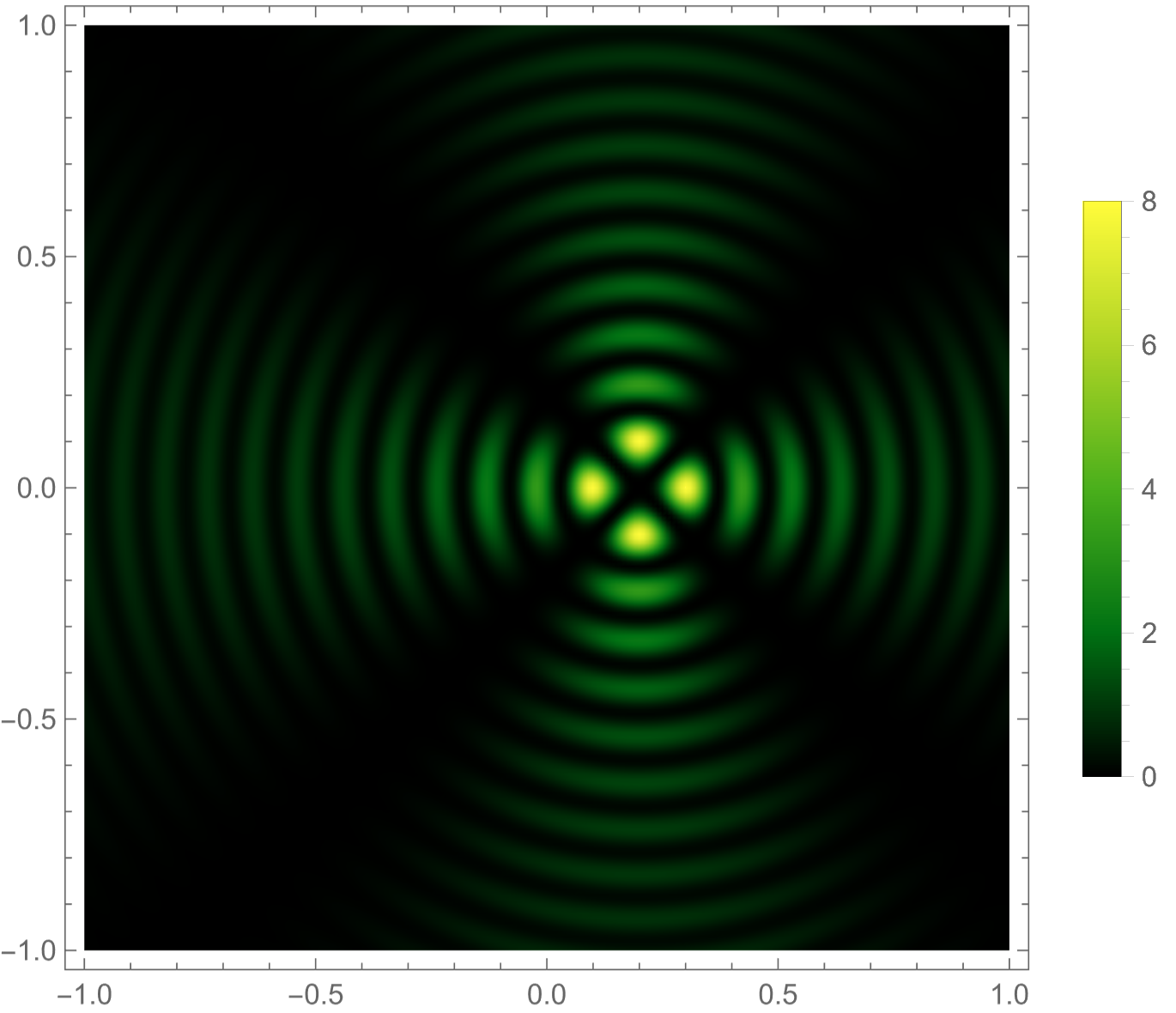}~\includegraphics[width=0.5\linewidth]{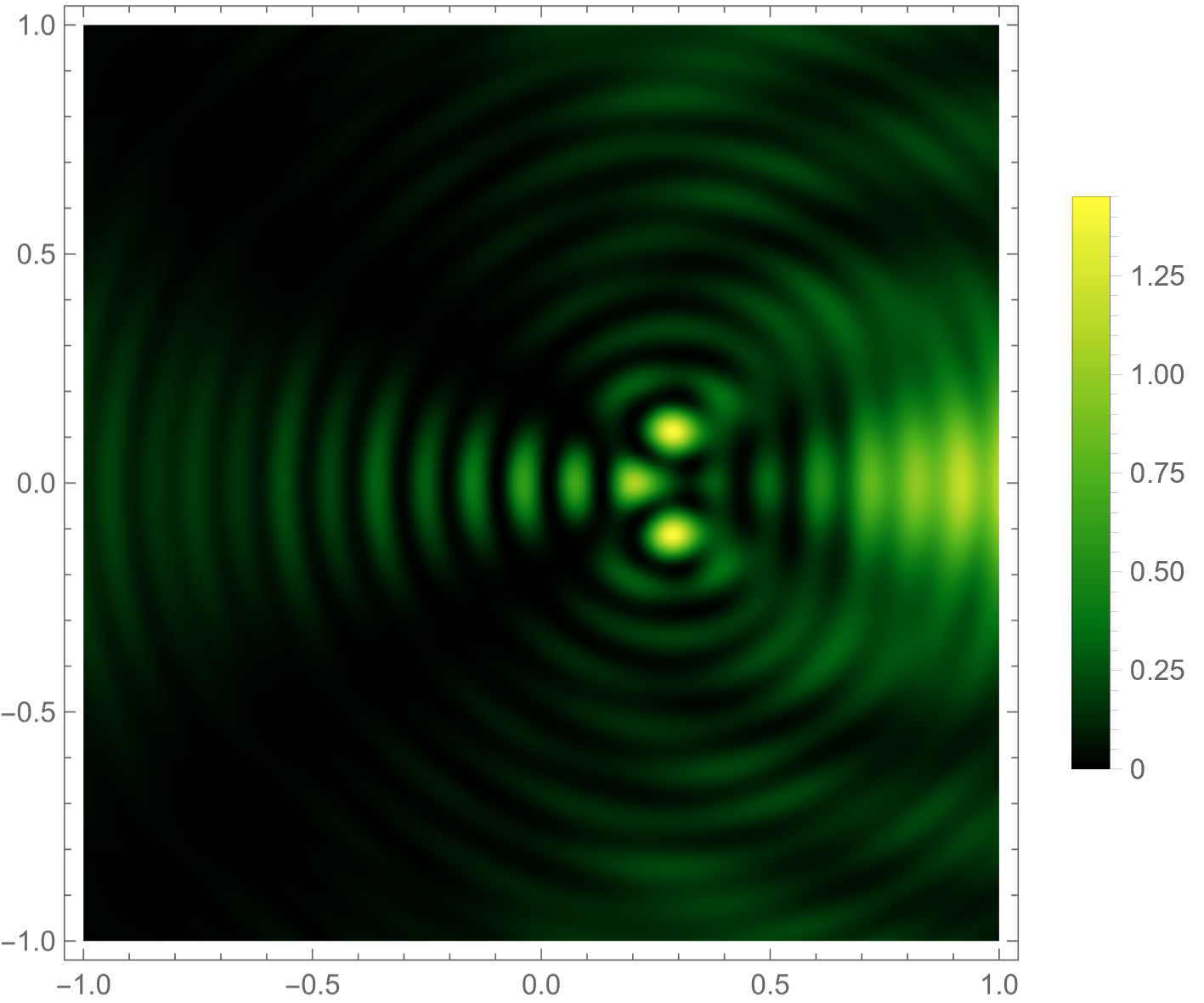}
    \caption{The second order correction in $\alpha$ to the shifted nonrotating interference pattern (left) and its difference with respect to the exact Kirchhoff-Fresnel integral (right), with the latter capturing the higher-order contributions given by $\mathrm{O}(\alpha^3)$. We used $\bm{\alpha} =(0.2,0)$ and $w=10$ in the first row and $w=30$ in the second. The brightness of the image is also shown for comparison.    }
    \label{fig:second-order_higher-order}
\end{figure}

\end{widetext} 
We did not find a closed-form expression for the remaining integral over $\theta$ in $\Delta \Psi$. However, as this integral is well-behaved, and runs over a finite domain, we evaluate it numerically. In Fig.~\ref{fig:second-order_higher-order} we show the interference pattern created by the second order correction term in $\alpha$ in Eq.~\eqref{eq:delta-psi}  and its difference with the right column of Fig.~\ref{fig:comparison} for $w=10$ (first row) and $w=30$ (second row). The right column of Fig.~\ref{fig:second-order_higher-order} represents the higher-order corrections not captured by Eq.~\eqref{eq:delta-psi}. The second-order correction (left column in Fig.~\ref{fig:second-order_higher-order}) already encodes the ``lighthouse'' effect that we observed in the right column of Fig.~\ref{fig:comparison}.
This shows that using the shifted non-rotating time delay leads to a loss of structure in the interference pattern, and the first statement above is incorrect.

Notice that the intensity of the second-order correction (left column in Fig.~\ref{fig:second-order_higher-order}) is comparable (for $w=10$) or even higher (for $w=30$) to the magnitude of $\alpha \approx 0.2$. Therefore, this difference is not negligible, which also invalidates the second statement above. Also note that the intensity of the higher order correction depicted in the right column of Fig.~\ref{fig:second-order_higher-order} is one order of magnitude smaller than the second order effect in $\alpha$ (left column). This is expected and hints at the possibility of approximating $\sum_{n=2}^N \frac{\alpha^n \cos n \bar{\theta}}{n x^n}$ by a finite number of terms $N>2$. As seen by looking at the intensity bars in Fig.~\ref{fig:second-order_higher-order}, for higher frequencies, $N$ needs to be larger to capture the structure of the interference pattern up to order $\alpha$.
However, we do not see any advantage in using Eq.~\eqref{eq:time-delay-approximation:1} truncated up to a certain $N$ rather than simply using Eq.~\eqref{eq:time-delay-dimensionless} since we have explicitly shown that in both cases, the $N=2$ correction \emph{must} be included, thereby already breaking the degeneracy of the caustics.

\section{Discussion}
We have calculated the interference pattern of a rotating lens in the small spin approximation using the Lense-Thirring metric. Despite previous claims in the literature, the resulting interference pattern is not merely a translated version of that produced by a non-rotating lens of equivalent mass; rather, it exhibits distinctive characteristics.  A comparative analysis of the interference patterns generated by the rotating lens and the displaced non-rotating lens reveals a prominent feature that bears resemblance to the light emitted from a lighthouse (refer to Fig.~\ref{fig:comparison}). 
If these features can be observed, this would be a new and independent methodology to measure the spin of rotating objects,  a task that is generally considered to be challenging.

In strong gravitational lensing, the lens bends the paths of the radiation and multiple images, arcs, or Einstein rings appear. In weak gravitational lensing, the deflection due to the presence of a lens is impossible to detect from a single background source. The presence of the foreground lens mass has to be statistically identified through the systematic alignment of multiple background sources around the lensing mass. In the present paper, we consider systems for which radiation experiences a strong deflection but for which the different images are not resolved. Rather, we observe the interference of the different images. This is known as gravitational microlensing. Not unlike Young's double slit experiment, the complementarity principle indicates that when we can identify the path of the rays, the interference pattern disappears. As a result, gravitational microlensing is often washed out. However, it is relevant for coherent radiation with long wavelengths, like the recently observed fast radio bursts and gravitational waves.

The interference patterns presented in this paper are not directly observed in the sky given that we do not have access to the entire image in the image plane. Rather, the effect of a gravitational microlens is a transient astronomical event. As the relative alignment of the source, lens and observer changes, the source's apparent brightness changes. The lens system traces a line in the presented interference pattern observed as fluctuations in intensity. The presented interference patterns can thus be interpreted as a collection of waveforms, where the caustics mark the most striking features. A frame-dragging detection by a microlens would thus require one to fit the presented model of the interference fringes to the light curve. 

Note that our results indicate that frame-dragging could in principle also be observed in strong gravitational lensing, through the relative positions of the images in the sky (as one should consider the time delay \eqref{eq:time-delay-dimensionless} rather than the approximation \eqref{eq:approx}).

The discussion in this paper is tailored to the language of wave optics as observable in electromagnetic radiation. However, wave optics may also be observable in gravitational wave observations and the relevant mathematics is analogous. The observable in this context is the waveform. The lensed waveform is computed from the unlensed one by convolving it with the Kirchoff-Fresnel integral in the frequency domain. 
The prospect of observing wave optics with gravitational waves at low frequencies with the Laser Interferometer Space Antenna (LISA) is particularly promising. In the point mass approximation for the lens, lens masses as ``small'' as $\sim 10^3 M_\odot$ can be inferred from the observed gravitational wave from a massive black hole binary with a total mass of $\sim 10^6 M_\odot$\cite{Caliskan:2022hbu}. It would be interesting to investigate the observability of a spinning lens using gravitational waves.

The foundations laid down in this article provide a stepping stone for a full derivation of a rotating black hole as lens. This would require a derivation of the time delay function using the Kerr metric instead of its small spin limit and take strong field effects into account (such as the possibility of rays disappearing into the black hole and rays going around the black hole multiple times before reemerging).
\vspace{0.5cm}

\section*{Acknowledgments}
BB is grateful to Neal Dalal and Suvendu Giri for useful discussions. JF is grateful for the discussions with Ue-Li Pen.

BB would like to thank the Perimeter Institute for Theoretical Physics, where the final phases of this work were completed.
Research at Perimeter Institute is
supported by the Government of Canada through the Department of Innovation, Science and Economic Development
and by the Province of Ontario through the Ministry of Colleges and Universities.

The work of JF is supported by the STFC Consolidated Grant ‘Particle Physics at the Higgs Centre,’ and, respectively, by a Higgs Fellowship and the Higgs Chair of Theoretical Physics at the University of Edinburgh.

For the purpose of open access, the author has applied a Creative Commons Attribution (CC BY) license to any Author Accepted Manuscript version arising from this submission.

\bibliographystyle{apsrev4-1}
\bibliography{library.bib}

\begin{thebibliography}{35}%
\makeatletter
\providecommand \@ifxundefined [1]{%
 \@ifx{#1\undefined}
}%
\providecommand \@ifnum [1]{%
 \ifnum #1\expandafter \@firstoftwo
 \else \expandafter \@secondoftwo
 \fi
}%
\providecommand \@ifx [1]{%
 \ifx #1\expandafter \@firstoftwo
 \else \expandafter \@secondoftwo
 \fi
}%
\providecommand \natexlab [1]{#1}%
\providecommand \enquote  [1]{``#1''}%
\providecommand \bibnamefont  [1]{#1}%
\providecommand \bibfnamefont [1]{#1}%
\providecommand \citenamefont [1]{#1}%
\providecommand \href@noop [0]{\@secondoftwo}%
\providecommand \href [0]{\begingroup \@sanitize@url \@href}%
\providecommand \@href[1]{\@@startlink{#1}\@@href}%
\providecommand \@@href[1]{\endgroup#1\@@endlink}%
\providecommand \@sanitize@url [0]{\catcode `\\12\catcode `\$12\catcode `\&12\catcode `\#12\catcode `\^12\catcode `\_12\catcode `\%12\relax}%
\providecommand \@@startlink[1]{}%
\providecommand \@@endlink[0]{}%
\providecommand \url  [0]{\begingroup\@sanitize@url \@url }%
\providecommand \@url [1]{\endgroup\@href {#1}{\urlprefix }}%
\providecommand \urlprefix  [0]{URL }%
\providecommand \Eprint [0]{\href }%
\providecommand \doibase [0]{http://dx.doi.org/}%
\providecommand \selectlanguage [0]{\@gobble}%
\providecommand \bibinfo  [0]{\@secondoftwo}%
\providecommand \bibfield  [0]{\@secondoftwo}%
\providecommand \translation [1]{[#1]}%
\providecommand \BibitemOpen [0]{}%
\providecommand \bibitemStop [0]{}%
\providecommand \bibitemNoStop [0]{.\EOS\space}%
\providecommand \EOS [0]{\spacefactor3000\relax}%
\providecommand \BibitemShut  [1]{\csname bibitem#1\endcsname}%
\let\auto@bib@innerbib\@empty
\bibitem [{\citenamefont {{Baraldo}}\ \emph {et~al.}(1999)\citenamefont {{Baraldo}}, \citenamefont {{Hosoya}},\ and\ \citenamefont {{Nakamura}}}]{Baraldo:1999}%
  \BibitemOpen
  \bibfield  {author} {\bibinfo {author} {\bibfnamefont {C.}~\bibnamefont {{Baraldo}}}, \bibinfo {author} {\bibfnamefont {A.}~\bibnamefont {{Hosoya}}}, \ and\ \bibinfo {author} {\bibfnamefont {T.~T.}\ \bibnamefont {{Nakamura}}},\ }\href {\doibase 10.1103/PhysRevD.59.083001} {\bibfield  {journal} {\bibinfo  {journal} {\prd}\ }\textbf {\bibinfo {volume} {59}},\ \bibinfo {eid} {083001} (\bibinfo {year} {1999})}\BibitemShut {NoStop}%
\bibitem [{\citenamefont {{Petroff}}\ \emph {et~al.}(2019)\citenamefont {{Petroff}}, \citenamefont {{Hessels}},\ and\ \citenamefont {{Lorimer}}}]{Petroff:2019}%
  \BibitemOpen
  \bibfield  {author} {\bibinfo {author} {\bibfnamefont {E.}~\bibnamefont {{Petroff}}}, \bibinfo {author} {\bibfnamefont {J.~W.~T.}\ \bibnamefont {{Hessels}}}, \ and\ \bibinfo {author} {\bibfnamefont {D.~R.}\ \bibnamefont {{Lorimer}}},\ }\href {\doibase 10.1007/s00159-019-0116-6} {\bibfield  {journal} {\bibinfo  {journal} {\aapr}\ }\textbf {\bibinfo {volume} {27}},\ \bibinfo {eid} {4} (\bibinfo {year} {2019})},\ \Eprint {http://arxiv.org/abs/1904.07947} {arXiv:1904.07947 [astro-ph.HE]} \BibitemShut {NoStop}%
\bibitem [{\citenamefont {{Abbott}}(2016)}]{Abott:2016}%
  \BibitemOpen
  \bibfield  {author} {\bibinfo {author} {\bibfnamefont {e.~a.}\ \bibnamefont {{Abbott}}, \bibfnamefont {B.~P.}},\ }\href {\doibase 10.1103/PhysRevLett.116.061102} {\bibfield  {journal} {\bibinfo  {journal} {\prl}\ }\textbf {\bibinfo {volume} {116}},\ \bibinfo {eid} {061102} (\bibinfo {year} {2016})},\ \Eprint {http://arxiv.org/abs/1602.03837} {arXiv:1602.03837 [gr-qc]} \BibitemShut {NoStop}%
\bibitem [{\citenamefont {{Nakamura}}\ and\ \citenamefont {{Deguchi}}(1999)}]{Nakamura:1999}%
  \BibitemOpen
  \bibfield  {author} {\bibinfo {author} {\bibfnamefont {T.~T.}\ \bibnamefont {{Nakamura}}}\ and\ \bibinfo {author} {\bibfnamefont {S.}~\bibnamefont {{Deguchi}}},\ }\href {\doibase 10.1143/PTPS.133.137} {\bibfield  {journal} {\bibinfo  {journal} {Progress of Theoretical Physics Supplement}\ }\textbf {\bibinfo {volume} {133}},\ \bibinfo {pages} {137} (\bibinfo {year} {1999})}\BibitemShut {NoStop}%
\bibitem [{\citenamefont {{Grillo}}\ and\ \citenamefont {{Cordes}}(2018)}]{Grillo:2018}%
  \BibitemOpen
  \bibfield  {author} {\bibinfo {author} {\bibfnamefont {G.}~\bibnamefont {{Grillo}}}\ and\ \bibinfo {author} {\bibfnamefont {J.}~\bibnamefont {{Cordes}}},\ }\href {\doibase 10.48550/arXiv.1810.09058} {\bibfield  {journal} {\bibinfo  {journal} {arXiv e-prints}\ ,\ \bibinfo {eid} {arXiv:1810.09058}} (\bibinfo {year} {2018})},\ \Eprint {http://arxiv.org/abs/1810.09058} {arXiv:1810.09058 [astro-ph.CO]} \BibitemShut {NoStop}%
\bibitem [{\citenamefont {{Feldbrugge}}\ and\ \citenamefont {{Turok}}(2020)}]{Feldbrugge:2020}%
  \BibitemOpen
  \bibfield  {author} {\bibinfo {author} {\bibfnamefont {J.}~\bibnamefont {{Feldbrugge}}}\ and\ \bibinfo {author} {\bibfnamefont {N.}~\bibnamefont {{Turok}}},\ }\href {\doibase 10.48550/arXiv.2008.01154} {\bibfield  {journal} {\bibinfo  {journal} {arXiv e-prints}\ ,\ \bibinfo {eid} {arXiv:2008.01154}} (\bibinfo {year} {2020})},\ \Eprint {http://arxiv.org/abs/2008.01154} {arXiv:2008.01154 [gr-qc]} \BibitemShut {NoStop}%
\bibitem [{\citenamefont {{Tambalo}}\ \emph {et~al.}(2023)\citenamefont {{Tambalo}}, \citenamefont {{Zumalac{\'a}rregui}}, \citenamefont {{Dai}},\ and\ \citenamefont {{Cheung}}}]{Tambalo:2023}%
  \BibitemOpen
  \bibfield  {author} {\bibinfo {author} {\bibfnamefont {G.}~\bibnamefont {{Tambalo}}}, \bibinfo {author} {\bibfnamefont {M.}~\bibnamefont {{Zumalac{\'a}rregui}}}, \bibinfo {author} {\bibfnamefont {L.}~\bibnamefont {{Dai}}}, \ and\ \bibinfo {author} {\bibfnamefont {M.~H.-Y.}\ \bibnamefont {{Cheung}}},\ }\href {\doibase 10.1103/PhysRevD.108.043527} {\bibfield  {journal} {\bibinfo  {journal} {\prd}\ }\textbf {\bibinfo {volume} {108}},\ \bibinfo {eid} {043527} (\bibinfo {year} {2023})},\ \Eprint {http://arxiv.org/abs/2210.05658} {arXiv:2210.05658 [gr-qc]} \BibitemShut {NoStop}%
\bibitem [{\citenamefont {{Feldbrugge}}\ \emph {et~al.}(2023)\citenamefont {{Feldbrugge}}, \citenamefont {{Pen}},\ and\ \citenamefont {{Turok}}}]{Feldbrugge:2023}%
  \BibitemOpen
  \bibfield  {author} {\bibinfo {author} {\bibfnamefont {J.}~\bibnamefont {{Feldbrugge}}}, \bibinfo {author} {\bibfnamefont {U.-L.}\ \bibnamefont {{Pen}}}, \ and\ \bibinfo {author} {\bibfnamefont {N.}~\bibnamefont {{Turok}}},\ }\href {\doibase 10.1016/j.aop.2023.169255} {\bibfield  {journal} {\bibinfo  {journal} {Annals of Physics}\ }\textbf {\bibinfo {volume} {451}},\ \bibinfo {eid} {169255} (\bibinfo {year} {2023})}\BibitemShut {NoStop}%
\bibitem [{\citenamefont {{Feldbrugge}}(2023)}]{Feldbrugge:2023b}%
  \BibitemOpen
  \bibfield  {author} {\bibinfo {author} {\bibfnamefont {J.}~\bibnamefont {{Feldbrugge}}},\ }\href {\doibase 10.1093/mnras/stad349} {\bibfield  {journal} {\bibinfo  {journal} {\mnras}\ }\textbf {\bibinfo {volume} {520}},\ \bibinfo {pages} {2995} (\bibinfo {year} {2023})},\ \Eprint {http://arxiv.org/abs/2010.03089} {arXiv:2010.03089 [astro-ph.CO]} \BibitemShut {NoStop}%
\bibitem [{\citenamefont {{Jow}}\ \emph {et~al.}(2023)\citenamefont {{Jow}}, \citenamefont {{Pen}},\ and\ \citenamefont {{Feldbrugge}}}]{Jow:2023}%
  \BibitemOpen
  \bibfield  {author} {\bibinfo {author} {\bibfnamefont {D.~L.}\ \bibnamefont {{Jow}}}, \bibinfo {author} {\bibfnamefont {U.-L.}\ \bibnamefont {{Pen}}}, \ and\ \bibinfo {author} {\bibfnamefont {J.}~\bibnamefont {{Feldbrugge}}},\ }\href {\doibase 10.1093/mnras/stad2332} {\bibfield  {journal} {\bibinfo  {journal} {\mnras}\ }\textbf {\bibinfo {volume} {525}},\ \bibinfo {pages} {2107} (\bibinfo {year} {2023})},\ \Eprint {http://arxiv.org/abs/2204.12004} {arXiv:2204.12004 [astro-ph.HE]} \BibitemShut {NoStop}%
\bibitem [{\citenamefont {{Braga}}\ \emph {et~al.}(2024)\citenamefont {{Braga}}, \citenamefont {{Garoffolo}}, \citenamefont {{Ricciardone}}, \citenamefont {{Bartolo}},\ and\ \citenamefont {{Matarrese}}}]{Braga:2024}%
  \BibitemOpen
  \bibfield  {author} {\bibinfo {author} {\bibfnamefont {G.}~\bibnamefont {{Braga}}}, \bibinfo {author} {\bibfnamefont {A.}~\bibnamefont {{Garoffolo}}}, \bibinfo {author} {\bibfnamefont {A.}~\bibnamefont {{Ricciardone}}}, \bibinfo {author} {\bibfnamefont {N.}~\bibnamefont {{Bartolo}}}, \ and\ \bibinfo {author} {\bibfnamefont {S.}~\bibnamefont {{Matarrese}}},\ }\href {\doibase 10.48550/arXiv.2405.20208} {\bibfield  {journal} {\bibinfo  {journal} {arXiv e-prints}\ ,\ \bibinfo {eid} {arXiv:2405.20208}} (\bibinfo {year} {2024})},\ \Eprint {http://arxiv.org/abs/2405.20208} {arXiv:2405.20208 [astro-ph.CO]} \BibitemShut {NoStop}%
\bibitem [{\citenamefont {{Thirring}}(1918)}]{Thirring:1918}%
  \BibitemOpen
  \bibfield  {author} {\bibinfo {author} {\bibfnamefont {H.}~\bibnamefont {{Thirring}}},\ }\href@noop {} {\bibfield  {journal} {\bibinfo  {journal} {Physikalische Zeitschrift}\ }\textbf {\bibinfo {volume} {19}},\ \bibinfo {pages} {33} (\bibinfo {year} {1918})}\BibitemShut {NoStop}%
\bibitem [{\citenamefont {{Lense}}\ and\ \citenamefont {{Thirring}}(1918)}]{Lense:1918}%
  \BibitemOpen
  \bibfield  {author} {\bibinfo {author} {\bibfnamefont {J.}~\bibnamefont {{Lense}}}\ and\ \bibinfo {author} {\bibfnamefont {H.}~\bibnamefont {{Thirring}}},\ }\href@noop {} {\bibfield  {journal} {\bibinfo  {journal} {Physikalische Zeitschrift}\ }\textbf {\bibinfo {volume} {19}},\ \bibinfo {pages} {156} (\bibinfo {year} {1918})}\BibitemShut {NoStop}%
\bibitem [{\citenamefont {{Miller-Jones}}\ \emph {et~al.}(2019)\citenamefont {{Miller-Jones}}, \citenamefont {{Tetarenko}}, \citenamefont {{Sivakoff}}, \citenamefont {{Middleton}}, \citenamefont {{Altamirano}}, \citenamefont {{Anderson}}, \citenamefont {{Belloni}}, \citenamefont {{Fender}}, \citenamefont {{Jonker}}, \citenamefont {{K{\"o}rding}}, \citenamefont {{Krimm}}, \citenamefont {{Maitra}}, \citenamefont {{Markoff}}, \citenamefont {{Migliari}}, \citenamefont {{Mooley}}, \citenamefont {{Rupen}}, \citenamefont {{Russell}}, \citenamefont {{Russell}}, \citenamefont {{Sarazin}}, \citenamefont {{Soria}},\ and\ \citenamefont {{Tudose}}}]{Miller-Jones:2019}%
  \BibitemOpen
  \bibfield  {author} {\bibinfo {author} {\bibfnamefont {J.~C.~A.}\ \bibnamefont {{Miller-Jones}}}, \bibinfo {author} {\bibfnamefont {A.~J.}\ \bibnamefont {{Tetarenko}}}, \bibinfo {author} {\bibfnamefont {G.~R.}\ \bibnamefont {{Sivakoff}}}, \bibinfo {author} {\bibfnamefont {M.~J.}\ \bibnamefont {{Middleton}}}, \bibinfo {author} {\bibfnamefont {D.}~\bibnamefont {{Altamirano}}}, \bibinfo {author} {\bibfnamefont {G.~E.}\ \bibnamefont {{Anderson}}}, \bibinfo {author} {\bibfnamefont {T.~M.}\ \bibnamefont {{Belloni}}}, \bibinfo {author} {\bibfnamefont {R.~P.}\ \bibnamefont {{Fender}}}, \bibinfo {author} {\bibfnamefont {P.~G.}\ \bibnamefont {{Jonker}}}, \bibinfo {author} {\bibfnamefont {E.~G.}\ \bibnamefont {{K{\"o}rding}}}, \bibinfo {author} {\bibfnamefont {H.~A.}\ \bibnamefont {{Krimm}}}, \bibinfo {author} {\bibfnamefont {D.}~\bibnamefont {{Maitra}}}, \bibinfo {author} {\bibfnamefont {S.}~\bibnamefont {{Markoff}}}, \bibinfo {author} {\bibfnamefont {S.}~\bibnamefont {{Migliari}}}, \bibinfo {author} {\bibfnamefont
  {K.~P.}\ \bibnamefont {{Mooley}}}, \bibinfo {author} {\bibfnamefont {M.~P.}\ \bibnamefont {{Rupen}}}, \bibinfo {author} {\bibfnamefont {D.~M.}\ \bibnamefont {{Russell}}}, \bibinfo {author} {\bibfnamefont {T.~D.}\ \bibnamefont {{Russell}}}, \bibinfo {author} {\bibfnamefont {C.~L.}\ \bibnamefont {{Sarazin}}}, \bibinfo {author} {\bibfnamefont {R.}~\bibnamefont {{Soria}}}, \ and\ \bibinfo {author} {\bibfnamefont {V.}~\bibnamefont {{Tudose}}},\ }\href {\doibase 10.1038/s41586-019-1152-0} {\bibfield  {journal} {\bibinfo  {journal} {\nat}\ }\textbf {\bibinfo {volume} {569}},\ \bibinfo {pages} {374} (\bibinfo {year} {2019})},\ \Eprint {http://arxiv.org/abs/1906.05400} {arXiv:1906.05400 [astro-ph.HE]} \BibitemShut {NoStop}%
\bibitem [{\citenamefont {{Venkatraman Krishnan}}\ \emph {et~al.}(2020)\citenamefont {{Venkatraman Krishnan}}, \citenamefont {{Bailes}}, \citenamefont {{van Straten}}, \citenamefont {{Wex}}, \citenamefont {{Freire}}, \citenamefont {{Keane}}, \citenamefont {{Tauris}}, \citenamefont {{Rosado}}, \citenamefont {{Bhat}}, \citenamefont {{Flynn}}, \citenamefont {{Jameson}},\ and\ \citenamefont {{Os{\l}owski}}}]{Venkatraman:2020}%
  \BibitemOpen
  \bibfield  {author} {\bibinfo {author} {\bibfnamefont {V.}~\bibnamefont {{Venkatraman Krishnan}}}, \bibinfo {author} {\bibfnamefont {M.}~\bibnamefont {{Bailes}}}, \bibinfo {author} {\bibfnamefont {W.}~\bibnamefont {{van Straten}}}, \bibinfo {author} {\bibfnamefont {N.}~\bibnamefont {{Wex}}}, \bibinfo {author} {\bibfnamefont {P.~C.~C.}\ \bibnamefont {{Freire}}}, \bibinfo {author} {\bibfnamefont {E.~F.}\ \bibnamefont {{Keane}}}, \bibinfo {author} {\bibfnamefont {T.~M.}\ \bibnamefont {{Tauris}}}, \bibinfo {author} {\bibfnamefont {P.~A.}\ \bibnamefont {{Rosado}}}, \bibinfo {author} {\bibfnamefont {N.~D.~R.}\ \bibnamefont {{Bhat}}}, \bibinfo {author} {\bibfnamefont {C.}~\bibnamefont {{Flynn}}}, \bibinfo {author} {\bibfnamefont {A.}~\bibnamefont {{Jameson}}}, \ and\ \bibinfo {author} {\bibfnamefont {S.}~\bibnamefont {{Os{\l}owski}}},\ }\href {\doibase 10.1126/science.aax7007} {\bibfield  {journal} {\bibinfo  {journal} {Science}\ }\textbf {\bibinfo {volume} {367}},\ \bibinfo {pages} {577} (\bibinfo {year}
  {2020})},\ \Eprint {http://arxiv.org/abs/2001.11405} {arXiv:2001.11405 [astro-ph.HE]} \BibitemShut {NoStop}%
\bibitem [{\citenamefont {{Grould}}\ \emph {et~al.}(2017)\citenamefont {{Grould}}, \citenamefont {{Vincent}}, \citenamefont {{Paumard}},\ and\ \citenamefont {{Perrin}}}]{Grould:2017}%
  \BibitemOpen
  \bibfield  {author} {\bibinfo {author} {\bibfnamefont {M.}~\bibnamefont {{Grould}}}, \bibinfo {author} {\bibfnamefont {F.~H.}\ \bibnamefont {{Vincent}}}, \bibinfo {author} {\bibfnamefont {T.}~\bibnamefont {{Paumard}}}, \ and\ \bibinfo {author} {\bibfnamefont {G.}~\bibnamefont {{Perrin}}},\ }in\ \href {\doibase 10.1017/S174392131601245X} {\emph {\bibinfo {booktitle} {The Multi-Messenger Astrophysics of the Galactic Centre}}},\ \bibinfo {series} {IAU Symposium}, Vol.\ \bibinfo {volume} {322},\ \bibinfo {editor} {edited by\ \bibinfo {editor} {\bibfnamefont {R.~M.}\ \bibnamefont {{Crocker}}}, \bibinfo {editor} {\bibfnamefont {S.~N.}\ \bibnamefont {{Longmore}}}, \ and\ \bibinfo {editor} {\bibfnamefont {G.~V.}\ \bibnamefont {{Bicknell}}}}\ (\bibinfo {year} {2017})\ pp.\ \bibinfo {pages} {25--30}\BibitemShut {NoStop}%
\bibitem [{\citenamefont {Ulmer}\ and\ \citenamefont {Goodman}(1995)}]{Ulmer:1994ij}%
  \BibitemOpen
  \bibfield  {author} {\bibinfo {author} {\bibfnamefont {A.}~\bibnamefont {Ulmer}}\ and\ \bibinfo {author} {\bibfnamefont {J.}~\bibnamefont {Goodman}},\ }\href {\doibase 10.1086/175422} {\bibfield  {journal} {\bibinfo  {journal} {Astrophys. J.}\ }\textbf {\bibinfo {volume} {442}},\ \bibinfo {pages} {67} (\bibinfo {year} {1995})},\ \Eprint {http://arxiv.org/abs/astro-ph/9406042} {arXiv:astro-ph/9406042} \BibitemShut {NoStop}%
\bibitem [{\citenamefont {Takahashi}\ and\ \citenamefont {Nakamura}(2003)}]{Takahashi:2003ix}%
  \BibitemOpen
  \bibfield  {author} {\bibinfo {author} {\bibfnamefont {R.}~\bibnamefont {Takahashi}}\ and\ \bibinfo {author} {\bibfnamefont {T.}~\bibnamefont {Nakamura}},\ }\href {\doibase 10.1086/377430} {\bibfield  {journal} {\bibinfo  {journal} {Astrophys. J.}\ }\textbf {\bibinfo {volume} {595}},\ \bibinfo {pages} {1039} (\bibinfo {year} {2003})},\ \Eprint {http://arxiv.org/abs/astro-ph/0305055} {arXiv:astro-ph/0305055} \BibitemShut {NoStop}%
\bibitem [{\citenamefont {Matsunaga}\ and\ \citenamefont {Yamamoto}(2006)}]{Matsunaga:2006uc}%
  \BibitemOpen
  \bibfield  {author} {\bibinfo {author} {\bibfnamefont {N.}~\bibnamefont {Matsunaga}}\ and\ \bibinfo {author} {\bibfnamefont {K.}~\bibnamefont {Yamamoto}},\ }\href {\doibase 10.1088/1475-7516/2006/01/023} {\bibfield  {journal} {\bibinfo  {journal} {JCAP}\ }\textbf {\bibinfo {volume} {01}},\ \bibinfo {pages} {023} (\bibinfo {year} {2006})},\ \Eprint {http://arxiv.org/abs/astro-ph/0601701} {arXiv:astro-ph/0601701} \BibitemShut {NoStop}%
\bibitem [{\citenamefont {Asada}\ and\ \citenamefont {Kasai}(2000)}]{Asada:2000vn}%
  \BibitemOpen
  \bibfield  {author} {\bibinfo {author} {\bibfnamefont {H.}~\bibnamefont {Asada}}\ and\ \bibinfo {author} {\bibfnamefont {M.}~\bibnamefont {Kasai}},\ }\href {\doibase 10.1143/PTP.104.95} {\bibfield  {journal} {\bibinfo  {journal} {Prog. Theor. Phys.}\ }\textbf {\bibinfo {volume} {104}},\ \bibinfo {pages} {95} (\bibinfo {year} {2000})},\ \Eprint {http://arxiv.org/abs/astro-ph/0006157} {arXiv:astro-ph/0006157} \BibitemShut {NoStop}%
\bibitem [{\citenamefont {Sereno}(2002)}]{Sereno:2002tv}%
  \BibitemOpen
  \bibfield  {author} {\bibinfo {author} {\bibfnamefont {M.}~\bibnamefont {Sereno}},\ }\href {\doibase 10.1016/S0375-9601(02)01361-0} {\bibfield  {journal} {\bibinfo  {journal} {Phys. Lett. A}\ }\textbf {\bibinfo {volume} {305}},\ \bibinfo {pages} {7} (\bibinfo {year} {2002})},\ \Eprint {http://arxiv.org/abs/astro-ph/0209148} {arXiv:astro-ph/0209148} \BibitemShut {NoStop}%
\bibitem [{\citenamefont {{Ebrahimnejad Rahbari}}\ \emph {et~al.}(2005)\citenamefont {{Ebrahimnejad Rahbari}}, \citenamefont {{Nouri-Zonoz}},\ and\ \citenamefont {{Rahvar}}}]{Ebrahimnejad:2005}%
  \BibitemOpen
  \bibfield  {author} {\bibinfo {author} {\bibfnamefont {H.}~\bibnamefont {{Ebrahimnejad Rahbari}}}, \bibinfo {author} {\bibfnamefont {M.}~\bibnamefont {{Nouri-Zonoz}}}, \ and\ \bibinfo {author} {\bibfnamefont {S.}~\bibnamefont {{Rahvar}}},\ }\href {\doibase 10.48550/arXiv.astro-ph/0508477} {\bibfield  {journal} {\bibinfo  {journal} {arXiv e-prints}\ ,\ \bibinfo {eid} {astro-ph/0508477}} (\bibinfo {year} {2005})},\ \Eprint {http://arxiv.org/abs/astro-ph/0508477} {arXiv:astro-ph/0508477 [astro-ph]} \BibitemShut {NoStop}%
\bibitem [{\citenamefont {Feynman}\ and\ \citenamefont {Zee}(2006)}]{Feynman:2006}%
  \BibitemOpen
  \bibfield  {author} {\bibinfo {author} {\bibfnamefont {R.}~\bibnamefont {Feynman}}\ and\ \bibinfo {author} {\bibfnamefont {A.}~\bibnamefont {Zee}},\ }\href {https://books.google.co.uk/books?id=Uv-uxB0sRKEC} {\emph {\bibinfo {title} {QED: The Strange Theory of Light and Matter}}},\ Alix G. Mautner memorial lectures\ (\bibinfo  {publisher} {Princeton University Press},\ \bibinfo {year} {2006})\BibitemShut {NoStop}%
\bibitem [{\citenamefont {{Schneider}}\ \emph {et~al.}(1992)\citenamefont {{Schneider}}, \citenamefont {{Ehlers}},\ and\ \citenamefont {{Falco}}}]{Schneider:1992}%
  \BibitemOpen
  \bibfield  {author} {\bibinfo {author} {\bibfnamefont {P.}~\bibnamefont {{Schneider}}}, \bibinfo {author} {\bibfnamefont {J.}~\bibnamefont {{Ehlers}}}, \ and\ \bibinfo {author} {\bibfnamefont {E.~E.}\ \bibnamefont {{Falco}}},\ }\href {\doibase 10.1007/978-3-662-03758-4} {\emph {\bibinfo {title} {{Gravitational Lenses}}}}\ (\bibinfo {year} {1992})\BibitemShut {NoStop}%
\bibitem [{\citenamefont {{Ibanez}}(1983)}]{Ibanez:1983}%
  \BibitemOpen
  \bibfield  {author} {\bibinfo {author} {\bibfnamefont {J.}~\bibnamefont {{Ibanez}}},\ }\href@noop {} {\bibfield  {journal} {\bibinfo  {journal} {\aap}\ }\textbf {\bibinfo {volume} {124}},\ \bibinfo {pages} {175} (\bibinfo {year} {1983})}\BibitemShut {NoStop}%
\bibitem [{\citenamefont {{Dymnikova}}(1986)}]{Dymnikova:1986}%
  \BibitemOpen
  \bibfield  {author} {\bibinfo {author} {\bibfnamefont {I.~G.}\ \bibnamefont {{Dymnikova}}},\ }in\ \href@noop {} {\emph {\bibinfo {booktitle} {Relativity in Celestial Mechanics and Astrometry. High Precision Dynamical Theories and Observational Verifications}}},\ \bibinfo {series} {IAU Symposium}, Vol.\ \bibinfo {volume} {114},\ \bibinfo {editor} {edited by\ \bibinfo {editor} {\bibfnamefont {J.}~\bibnamefont {{Kovalevsky}}}\ and\ \bibinfo {editor} {\bibfnamefont {V.~A.}\ \bibnamefont {{Brumberg}}}}\ (\bibinfo {year} {1986})\ p.\ \bibinfo {pages} {411}\BibitemShut {NoStop}%
\bibitem [{\citenamefont {{Glicenstein}}(1999)}]{Glicenstein:1999}%
  \BibitemOpen
  \bibfield  {author} {\bibinfo {author} {\bibfnamefont {J.~F.}\ \bibnamefont {{Glicenstein}}},\ }\href@noop {} {\bibfield  {journal} {\bibinfo  {journal} {\aap}\ }\textbf {\bibinfo {volume} {343}},\ \bibinfo {pages} {1025} (\bibinfo {year} {1999})}\BibitemShut {NoStop}%
\bibitem [{\citenamefont {{Nye}}(1978)}]{Nye:1978}%
  \BibitemOpen
  \bibfield  {author} {\bibinfo {author} {\bibfnamefont {J.~F.}\ \bibnamefont {{Nye}}},\ }\href {\doibase 10.1098/rspa.1978.0090} {\bibfield  {journal} {\bibinfo  {journal} {Proceedings of the Royal Society of London Series A}\ }\textbf {\bibinfo {volume} {361}},\ \bibinfo {pages} {21} (\bibinfo {year} {1978})}\BibitemShut {NoStop}%
\bibitem [{\citenamefont {{Nye}}(1986)}]{Nye:1986}%
  \BibitemOpen
  \bibfield  {author} {\bibinfo {author} {\bibfnamefont {J.~F.}\ \bibnamefont {{Nye}}},\ }\href {\doibase 10.1098/rspa.1986.0001} {\bibfield  {journal} {\bibinfo  {journal} {Proceedings of the Royal Society of London Series A}\ }\textbf {\bibinfo {volume} {403}},\ \bibinfo {pages} {1} (\bibinfo {year} {1986})}\BibitemShut {NoStop}%
\bibitem [{\citenamefont {{Sereno}}(2003)}]{Sereno:2003}%
  \BibitemOpen
  \bibfield  {author} {\bibinfo {author} {\bibfnamefont {M.}~\bibnamefont {{Sereno}}},\ }\href {\doibase 10.1046/j.1365-8711.2003.06881.x} {\bibfield  {journal} {\bibinfo  {journal} {\mnras}\ }\textbf {\bibinfo {volume} {344}},\ \bibinfo {pages} {942} (\bibinfo {year} {2003})},\ \Eprint {http://arxiv.org/abs/astro-ph/0307243} {arXiv:astro-ph/0307243 [astro-ph]} \BibitemShut {NoStop}%
\bibitem [{\citenamefont {{Feldbrugge}}\ and\ \citenamefont {{Turok}}(2023)}]{Feldbrugge:2023c}%
  \BibitemOpen
  \bibfield  {author} {\bibinfo {author} {\bibfnamefont {J.}~\bibnamefont {{Feldbrugge}}}\ and\ \bibinfo {author} {\bibfnamefont {N.}~\bibnamefont {{Turok}}},\ }\href {\doibase 10.1016/j.aop.2023.169315} {\bibfield  {journal} {\bibinfo  {journal} {Annals of Physics}\ }\textbf {\bibinfo {volume} {454}},\ \bibinfo {eid} {169315} (\bibinfo {year} {2023})},\ \Eprint {http://arxiv.org/abs/2207.12798} {arXiv:2207.12798 [hep-th]} \BibitemShut {NoStop}%
\bibitem [{\citenamefont {\c{C}al\i{}\c{s}kan}\ \emph {et~al.}(2023)\citenamefont {\c{C}al\i{}\c{s}kan}, \citenamefont {Ji}, \citenamefont {Cotesta}, \citenamefont {Berti}, \citenamefont {Kamionkowski},\ and\ \citenamefont {Marsat}}]{Caliskan:2022hbu}%
  \BibitemOpen
  \bibfield  {author} {\bibinfo {author} {\bibfnamefont {M.}~\bibnamefont {\c{C}al\i{}\c{s}kan}}, \bibinfo {author} {\bibfnamefont {L.}~\bibnamefont {Ji}}, \bibinfo {author} {\bibfnamefont {R.}~\bibnamefont {Cotesta}}, \bibinfo {author} {\bibfnamefont {E.}~\bibnamefont {Berti}}, \bibinfo {author} {\bibfnamefont {M.}~\bibnamefont {Kamionkowski}}, \ and\ \bibinfo {author} {\bibfnamefont {S.}~\bibnamefont {Marsat}},\ }\href {\doibase 10.1103/PhysRevD.107.043029} {\bibfield  {journal} {\bibinfo  {journal} {Phys. Rev. D}\ }\textbf {\bibinfo {volume} {107}},\ \bibinfo {pages} {043029} (\bibinfo {year} {2023})},\ \Eprint {http://arxiv.org/abs/2206.02803} {arXiv:2206.02803 [astro-ph.CO]} \BibitemShut {NoStop}%
\bibitem [{\citenamefont {Thorne}(1980)}]{Thorne:1980ru}%
  \BibitemOpen
  \bibfield  {author} {\bibinfo {author} {\bibfnamefont {K.~S.}\ \bibnamefont {Thorne}},\ }\href {\doibase 10.1103/RevModPhys.52.299} {\bibfield  {journal} {\bibinfo  {journal} {Rev. Mod. Phys.}\ }\textbf {\bibinfo {volume} {52}},\ \bibinfo {pages} {299} (\bibinfo {year} {1980})}\BibitemShut {NoStop}%
\bibitem [{\citenamefont {Thorne}\ and\ \citenamefont {Blandford}(2017)}]{thorne2017modern}%
  \BibitemOpen
  \bibfield  {author} {\bibinfo {author} {\bibfnamefont {K.}~\bibnamefont {Thorne}}\ and\ \bibinfo {author} {\bibfnamefont {R.}~\bibnamefont {Blandford}},\ }\href {https://books.google.nl/books?id=eGmYDwAAQBAJ} {\emph {\bibinfo {title} {Modern Classical Physics: Optics, Fluids, Plasmas, Elasticity, Relativity, and Statistical Physics}}}\ (\bibinfo  {publisher} {Princeton University Press},\ \bibinfo {year} {2017})\BibitemShut {NoStop}%
\bibitem [{\citenamefont {Robinson}(1975)}]{PhysRevLett.34.905}%
  \BibitemOpen
  \bibfield  {author} {\bibinfo {author} {\bibfnamefont {D.~C.}\ \bibnamefont {Robinson}},\ }\href {\doibase 10.1103/PhysRevLett.34.905} {\bibfield  {journal} {\bibinfo  {journal} {Phys. Rev. Lett.}\ }\textbf {\bibinfo {volume} {34}},\ \bibinfo {pages} {905} (\bibinfo {year} {1975})}\BibitemShut {NoStop}%
\end{thebibliography}%

\appendix
\section{Unfolding of the point lens}\label{ap:unfolding}
\begin{figure}
  \centering
  \includegraphics[width=\linewidth]{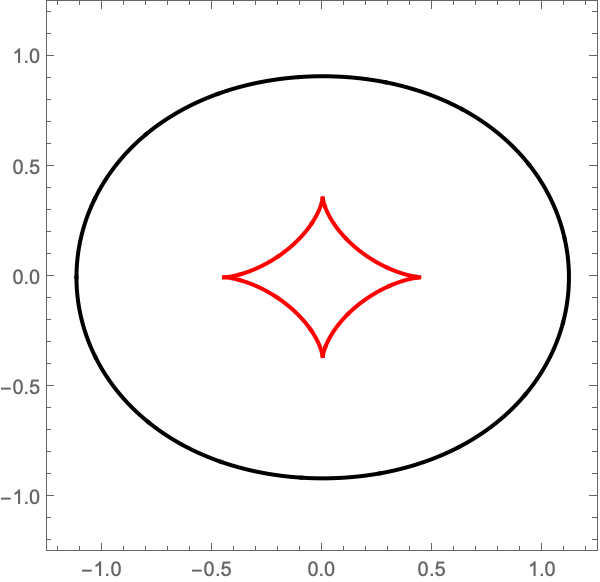}
  \caption{The critical curve (black) in the lens plane and caustic curve (red) in the image plane of a non-rotating point lens with a shear term superimposed on top of each other ($\gamma_1=0$ and $\gamma_2=0.2$).}\label{fig:astroid}
\end{figure}

The time delay of the non-rotating point lens $T(\bm{x},\bm{y}) = (\bm{x}-\bm{y})^2/2 - \log x$ has a one-dimensional critical curve $\mathcal{C}=\{\lVert\bm{x}\rVert = 1\}$ and a degenerate zero-dimensional caustic curve $\bm{\xi}(\mathcal{C}) = \{\bm{0}\}$. To study the unfolding of the caustic, let us perturb the time delay function,
\begin{align}
  T(\bm{x},\bm{y}) = \frac{(\bm{x}-\bm{y})^2}{2} - \log x +\delta T(\bm{x})
\end{align}
with a quadratic fluctuation $\delta T(\bm{x}) = c + \bm{x} \cdot \bm{\eta} + \frac{1}{2} \bm{x}^T \Sigma \bm{x}$ for a constant scalar $c$, $2$-vector $\bm{\eta}$ and symmetric $2\times 2$ matrix $\Sigma$. The constant term is irrelevant to the caustics. The linear term can be absorbed into the geometric contribution to the time delay function. For the quadratic term, we use the rotation symmetry of the unperturbed lens to rotate to the eigenframe of $\Sigma$,
\begin{align}
  \Sigma = \begin{pmatrix}\gamma_1 + \gamma_2 & 0\\ 0 & \gamma_1 - \gamma_2\end{pmatrix}\,.
\end{align} 
The generalized time delay thus assumes the form 
\begin{align}
  T(\bm{x},\bm{y})  
  \sim &
   \frac{(\bm{x}-(\bm{y}+\bm{\eta}))^2}{2} - \log x \nonumber\\
   & + \frac{\gamma_1}{2}(x_1^2+x_2^2) + \frac{\gamma_2}{2}(x_1^2- x_2^2)\,,
\end{align}
up to constant contributions. The corresponding lens map
\begin{align}
  \bm{\xi}(\bm{x}) = \bm{x} - \frac{\bm{x}}{x^2} + \bm{x} \cdot \begin{pmatrix}\gamma_1 + \gamma_2 \\ \gamma_1 - \gamma_2 \end{pmatrix}- \bm{\eta} 
\end{align}
shows that the linear term in the fluctuation $\delta T$ shifts the caustic by $\bm{y} \mapsto \bm{y} - \bm{\eta}$. The term $\gamma_1 x^2$ preserves the rotational symmetry of the lens and the degeneracy of the caustic curve. The critical curve is a circle with radius $1/\sqrt{1+ \gamma_1}$ and the caustic curve is the point $\{-\bm{\eta}\}$. The shear term $\frac{1}{2}\gamma_2 (x_1^2-x_2^2)$ breaks the rotation symmetry of the point lens. For small $\gamma_2$, the caustic is a one-dimensional astroid, consisting of four fold lines joined by four cusp points (see Fig.~\ref{fig:astroid}). This pattern is stable with respect to small perturbations.

\section{The small spin approximation}
\label{sec:small-spin}

In the main text, we use the small spin approximation, in which we assume that the modulus of the spin parameter $\alpha$ is small. However, we have not yet discussed the meaning of ``small'' in this context.  
Effectively, when deriving the time delay function, we used the Lense-Thirring metric, the weak field approximation of the Kerr metric in ACMC  (asymptotically Cartesian and mass centered) coordinates~\cite{Thorne:1980ru,thorne2017modern}, \textit{i.e.},  the Kerr metric expanded to first order in $r_s/r$ and the Kerr parameter $a$. The limitations of the weak field approximation are known: the Lense-Thirring metric is suitable to describe the spacetime surrounding stars, but does not apply to neutron stars or black holes~\cite{Glicenstein:1999}. The extension of the lens to a Kerr black hole will be discussed elsewhere. In this appendix, we will be concerned with the regularity of the next-order term in the small spin approximation. As it has extensively been discussed in the main text, one must be careful when taking the small spin limit since the terms accompanying the small spin parameter might diverge. Here, we compute the next order correction in the Kerr parameter $\sim a^2$ to the time delay and check that this term does not include singularities (apart from the physical singularity at $\bm{x} = 0$ discussed in Sec.~\ref{sec:time-delay}). 

\begin{figure}
    \centering
    \includegraphics[width=\linewidth]{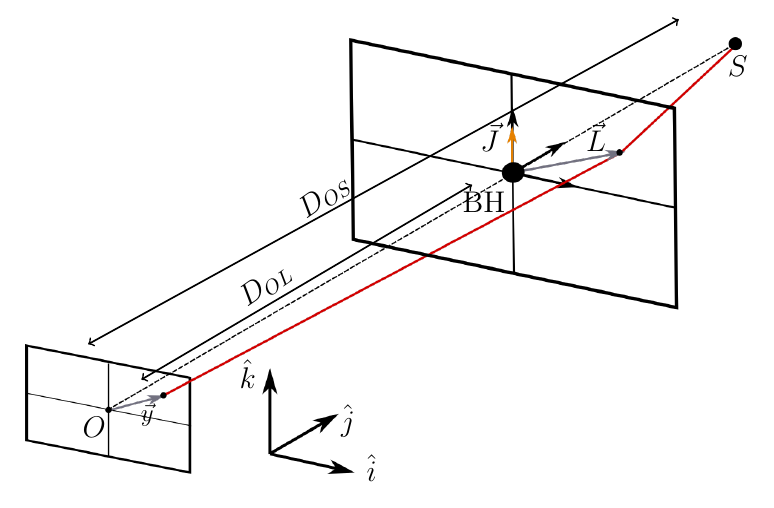}
    \caption{Schematics of a null ray being lensed by a slowly rotating Kerr black hole. The distance from the observer to the lens is $D_{OL}$, and the total distance from the source to the observer is $D_{OS}$. The coordinate system is placed on the black hole, with the spin vector pointing in the vertical direction. The ray intersects the lens plane at $\vec{L}$. }
    \label{fig:setup-kerr}
\end{figure}

The derivation presented here is for a simplified setup, which differs from the one detailed in the main text, but makes the derivation more transparent.  
The setup is represented in Fig.~\ref{fig:setup-kerr}, where the coordinate system is located on the black hole in the lens plane with the spin vector aligned with the $z$-axis. The black hole is located along the line of sight (represented by a dashed line) such that the $y$-axis lies along the line of sight and points towards the source. The lens is located at a distance $D_{OL}$ from the observer, while the source lies at a distance $D_{OS}$. These distances are considered to be much larger than the characteristic Schwarzschild radius $r_s=2M$ (in geometrized units $G=c=1$).
Given the set-up of the problem, in which the characteristic distances from the observer to the lens $D_{OL}$ and from the lens to the source $D_{LS}=D_{OS}-D_{OL}$ are much larger than the size of the black hole $r_s$, the signal from the source will propagate in a spacetime that is very similar to the Minkowski spacetime everywhere but close to the lens.

The Kerr metric in ACMC coordinates to quadratic order in the Kerr parameter $a$ and linear order in $r_s/r$ reads~\cite{Thorne:1980ru}
\begin{subequations}\label{eq:metric-kerr}
    \begin{align}
        g_{tt} &= -1+\frac{r_s}{r} -\frac{3 r_s a^2 \cos^2\theta}{2r^3}\\
        g_{t\phi} & = -\frac{a r_s \sin^2\theta}{r}\\
        g_{rr} & = 1+\frac{r_s}{r} -\frac{a^2}{r^2}-a^2\frac{r_s(4+\cos^2\theta)}{2r^3}\\
        g_{\theta\theta} & = r^2\left(1+\frac{a^2}{r^2}\right)\\
        g_{\phi\phi} & = r^2\sin^2\theta\left(1+\frac{a^2}{r^2} +\frac{r_s a^2 \sin^2\theta}{r^3}\right)\\
        g_{r\theta} & = -\frac{r_s a^2 \cos\theta \sin\theta}{r^2} \, .
    \end{align}
\end{subequations}
From this metric, it is straightforward to compute the null geodesic equation
\begin{subequations}\label{eq:geodesic-thorne}
    \begin{align}
        \dot{t}  = & E\left(1+\frac{r_s}{r}\right) -\frac{a J }{r^2} \frac{r_s}{r} -3 E\frac{a^2}{2r^2}\frac{r_s}{r} \cos^2\theta\\
        \dot{\phi}  = &\frac{J}{r^2\sin^2\theta} +\frac{a E}{r^2}\frac{r_s}{r} -\frac{a^2 J}{r^4}\left(\frac{1}{ \sin^2\theta}+\frac{r_s}{r}\right) \\
        \begin{split}
            0=& (r^2+a^2)\ddot{\theta}  +2r \dot{r}\dot{\theta}  -\frac{(r^2-a^2)J^2}{r^4\sin^2\theta} \cot\theta\\
            &-\frac{a^2 r_s}{4r^3} \sin2\theta (3E^2 -3\dot{r}^2 +2 r \ddot{r}) 
        \end{split}\\
        \begin{split}
             \dot{r}^2  & =  E^2 \left(1+\frac{a^2 (r+r_s\sin^2\theta )}{r^3}\right)  -\frac{2aJr_s E}{r^3}\\
             &-\frac{r-r_s}{r^3} (r^4 \dot{\theta}^2 +J^2 \csc^2\theta)\\
             &+\frac{a^2}{4r^5} \Big(6J^2 r_s +r^4 (3r_s-8r) \dot{\theta}^2 -\frac{6r_s J^2}{ \sin^2\theta}\\
             &-r_s r^3 \dot{\theta} (r\dot{\theta}\cos2\theta -4\dot{r} \sin 2\theta)\Big) \,,
        \end{split}
    \end{align}
\end{subequations}
where $E$ and $J$ are the energy and angular momentum of a test particle moving on a null geodesic. The dot indicates a derivative with respect to the proper time along the geodesic trajectory. Notice that up to order we are working in, that is, $\mathcal{O}(a^3\,,r_s a^2\,, r_s^2)$ the geodesic equation coincides with the usual geodesic equation of the Kerr metric in Boyer-Lindquist coordinates. 

The time delay is the proper time as measured by an observer at rest between two events: the emission of light from a faraway source and the reception of the ray. 
The relationship between the proper time of the observer and the coordinate time to first order in $r_s/D_{OL}$ is simply~\cite{thorne2017modern}
\begin{equation}
    \Delta\tau = \sqrt{-g_{tt}} \Delta t=\left(1-\frac{r_s}{2 D_{OL}} + \frac{3a^2 }{4D_{OL}^2} \frac{r_s}{D_{OL}}\cos^2\theta\right)\Delta t
\end{equation} where $\Delta t$ is the coordinate time it takes the ray to travel from the source to the observer after being deviated by the lens. Taking into account that we work in the thin lens approximation so that the distance $D_{OL}$ is large compared to $r_s$, we can approximate 
\begin{equation}
    \Delta \tau\approx \Delta t\,.
\end{equation} Hence, we define the time delay as 
\begin{equation}
    T = \Delta t|_{\text{source}}^{\text{observer}}
\end{equation} up to order $\mathcal{O}(a^3\,,r_s a^2\,, r_s^2\,, r_s/D_{OL})$, which using the geodesic equation Eq.~\eqref{eq:geodesic-thorne} reads
\begin{subequations}\label{eq:time-delay-kerr-geodesic}
    \begin{align}
         T &= \int_0^1 \dot{t} \, \d \tau \\
            &=  \int_0^1 \left(E\left(1+\frac{r_s}{r}\right) -\frac{a J }{r^2} \frac{r_s}{r} -3 E\frac{a^2}{2r^2}\frac{r_s}{r} \cos^2\theta\right)\d \tau  
    \end{align}
\end{subequations} where the integral over the proper time $\tau$ is taken along the geometric path (described in Fig.~\ref{fig:setup-kerr}) that the ray would follow in the absence of the lens. Namely, 
\begin{subequations}
    \begin{align}
        \dot{t} & = E\\
        \dot{\phi} & = \frac{J}{r^2\sin^2\theta} \left(1-\frac{a^2}{r^2}\right)\\
         0&= (r^2+a^2)\ddot{\theta}  +2r \dot{r}\dot{\theta}  -\frac{(r^2-a^2)J^2}{r^4\sin^2\theta} \cot\theta\\
         \dot{r}^2&=E^2 \left(1+\frac{a^2}{r^2}\right) -(2a^2 +r^2) \dot{\theta}^2-\frac{J^2\csc^2\theta}{r^2}\,.
    \end{align}
\end{subequations} This system of differential equations can be solved by realizing that the Kerr parameter only enters as a quadratic correction to the trajectory. Hence, we need to expand all of the relevant quantities as
\begin{subequations}\label{eq:trajectory}
    \begin{align}
        t(\tau) &= t_0(\tau) + a^2 t_2(\tau)\\
        r(\tau) &= r_0(\tau) + a^2 r_2(\tau)\\
        \theta(\tau) &= \theta_0(\tau) + a^2 \theta_2(\tau)\\
        \phi(\tau) &= \phi_0(\tau) + a^2 \phi_2(\tau)\\
        E & = E_0 + a^2 E_2\\
        J & = J_0 +a^2 J_2
    \end{align}
\end{subequations} The solution to the geodesic equation to zeroth order in the Kerr parameter is \begin{subequations}\label{eq:trajectory-0}
    \begin{align}
        r_0 (\tau) &= \sqrt{D_{LS}^2(\tau-1)^2 +L^2 \tau^2}\\
        \theta_0 (\tau) &= \arctan \left(\frac{\sqrt{D_{LS}^2 (\tau-1)^2 +L_1^2 \tau^2}}{L_2 \tau}\right)\\
        \phi_0 & = \arctan \left(\frac{D_{LS}(1-\tau)}{L_1 \tau}\right)\\
        J_0 & = -D_{LS} L_1\\
        E_0 & = \sqrt{D_{LS}^2 + L^2}
    \end{align}
\end{subequations} where we have used that 
the position of the source and the lens in Cartesian coordinates are $\bm{x} (\tau = 0) = (0\,,D_{LS}\,, 0)$ and $\bm{x} (\tau=1) = (L_1,0,L_2)$, and where we have defined 
$L = \lVert \bm{L}\rVert=\sqrt{L_1^2+L_2^2}$. This solution corresponds to a straight line in Minkowski space. For the correction of order $a^2$ we obtain
\begin{subequations}\label{eq:trajectory-2}
    \begin{align}
        r_2(\tau) &= \frac{\tau (\tau-1)}{\sqrt{D_{LS}^2 (\tau-1)^2 +L^2\tau^2}} \\
        \theta_2 (\tau) & = \frac{L_2 (L^2 -D_{LS}^2) \tau (1-\tau)^2}{2L^2 \sqrt{D_{LS}^2 (\tau-1)^2 +L_1^2\tau^2} (D_{LS}^2 (1-\tau)^2 +L^2\tau^2)} \\
        \phi_2 (\tau) & = \frac{L_1 (L^2-D_{LS}^2) \tau (1-\tau)}{2 (L^2) D_{LS}(D_{LS}^2(1-\tau)^2 +L_1^2 \tau^2)}\\
        J_2 & = -\frac{L_1 (L^2+D_{LS}^2)}{2D_{LS} L^2}\\
        E_2 & = \frac{1}{\sqrt{D_{LS}^2+L^2}}\,,
    \end{align}
\end{subequations} where we have fixed the integration constant by fixing the extremum of the path to its initial and final values according to Fig.~\ref{fig:setup-kerr}.

The time delay from the source to the lens plane is computed using Eq.~\eqref{eq:time-delay-kerr-geodesic} together with the path~\eqref{eq:trajectory}-\eqref{eq:trajectory-2}. 
Similar to the path and the constants of motion $E$ and $J$, we can expand the time delay function in terms of $a$
\begin{equation}\label{eq:time-delay-expansion-kerr-app}
    T=  T_0 +a  T_1+ a^2  T_2 +\mathcal{O}(a^3, r_s^2)\,,
\end{equation} where the time delay to zeroth order in the spin is
\begin{equation}\label{eq:time-delay-expansion-kerr-app:0}
     T_0 =\int_0^1 E_0 \left(1+\frac{r_s}{r_0}\right)\,  \d\tau\,,
\end{equation} to first-order 
\begin{equation} \label{eq:time-delay-expansion-kerr-app:1}
      T_1 =-J_0 r_s\int_0^1 \frac{1}{r_0^3}\, \d\tau\,,
\end{equation} and to second-order
\begin{equation}\label{eq:time-delay-expansion-kerr-app:2}
     T_2 =\int_0^1\left[ E_2 \left(1+\frac{r_s}{r_0}\right) -E_0 \frac{r_s}{r_0} \left(\frac{r_2}{r_0} +\frac{3}{2r_0^2} \cos^2\theta_0\right)\right] \,\d\tau\,.
\end{equation}The last term codifies the time delay due to the rotating background to the second order in the spin. Evaluating these expressions is cumbersome but otherwise straightforward using Eqs.~\eqref{eq:trajectory}-\eqref{eq:trajectory-2}. The first term yields
\begin{equation}
     T_0 = \sqrt{D_{LS}^2+L^2} +r_s \log\left(\frac{L^2+L\sqrt{D_{LS}^2+L^2}}{D_{LS}\sqrt{D_{LS}^2+L^2}-D_{LS}^2}\right)\,,
\end{equation} 
the linear term in the spin is 
\begin{equation}\label{eq:T1-SL}
     T_1 = \frac{L_1 r_s (D_{LS}+L)}{D_{LS} L^2}\,,
\end{equation} while the quadratic term in the spin is 
\begin{equation}
\begin{split}\label{eq:T2-SL}
     T_2 &= \frac{1}{\sqrt{D_{LS}^2+L^2}} \\
     &+r_s\frac{-2 D_{LS}^3L_2^2 +D_{LS}^2 L (L_1^2+L^2) +2 D_{LS} L_1^2 L^2 -L_2^2 L^3}{2D_{LS}^2L^4\sqrt{D_{LS}^2 + L^2}}\,.
\end{split} 
\end{equation} 

Next, we need to evaluate the time delay along the second half of the path $(r(\tau),\theta(\tau),\phi(\tau))$, which can be obtained by replacing $\tau\to 1-\tau$ and $D_{LS}\to-D_{OL}$ in Eqs.~\eqref{eq:trajectory}-\eqref{eq:trajectory-2} for an observer located at the line of sight. 
Solving the geodesic equation with these boundary conditions and evaluating Eqs.~\eqref{eq:time-delay-expansion-kerr-app}-\eqref{eq:time-delay-expansion-kerr-app:2} yields
\begin{equation}
     T_0 = \sqrt{D_{OL}^2+L^2} +r_s \log\left(\frac{D_{OL}^2+D_{OL}\sqrt{D_{OL}^2+L^2}}{L\sqrt{D_{OL}^2+L^2}-L^2}\right) \,.
\end{equation} for the zeroth order contribution to the time delay. The first and second-order contributions for the observer-lens path can be obtained by replacing $D_{LS}\to -D_{OL}$ in Eqs.~\eqref{eq:T1-SL} and~\eqref{eq:T2-SL}.

Combining the time delays for the paths between the source and lens and the lens and image plane, and expanding to first order in $1/D_{LS}$ and $1/D_{OL}$, and zeroth order in $r_s/D_{LS}$ and $r_s/D_{OL}$ yields
\begin{equation}\label{eq:time-delay-kerr-order2}
\begin{split}
     T&= D_{OS} +\frac{D_{OS}L^2}{2D_{LS}D_{OL}} -2r_s \log \left(\frac{L}{2\sqrt{D_{LS}D_{OL}}} \right) \\
      &+\frac{2 ar_s L_1}{L^2}-a^2  \frac{2L_2^2 r_s}{L^4}
\end{split}
\end{equation}
 
 Using an analogous normalization to the one discussed in the main text~\eqref{eq:normalization-x-y}, but taking into account that $\bm{L}$ is a \emph{distance} in the lens plane rather than an angle,
 \begin{equation}
     \bm{\alpha} = \frac{\bm{a}\times \bm{n}}{r_E} = -\frac{a}{r_E}(1\,,0)\,,\quad \bm{x} = \frac{\bm{L}}{r_E}\,,
 \end{equation} with the Einstein radius
 \begin{equation}
     r_E = \sqrt{2r_s \frac{D_{LS} D_{OL}}{D_{OS}}}\,,
 \end{equation} 
 and $n=(0,1,0)$, we obtain
 \begin{equation}\label{eq:time-delay-2nd order}
     T(x)= 2r_s \left(\phi_0 + \frac{\lVert\bm{x}-\bm{y}\rVert^2}{2} - \log 
    x -\frac{\alpha x_1}{x^2} -\frac{\alpha^2 x_2^2}{x^4}\right)
 \end{equation} where $x=\lVert \bm{x} \rVert$ is the norm of the vector $\bm{x}=(x_1,x_2)$ and we have taken into account the shift in the image plane given by $\bm{y}$ (see Fig.~\ref{fig:setup-kerr}).  The phase $\phi_0$ is the collection of the constant terms (independent of $\bm{L}$) in Eq.~\eqref{eq:time-delay-kerr-order2}. In covariant form Eq.~\eqref{eq:time-delay-2nd order} reads
 \begin{equation}\label{eq:time-delay-covariant}
    T(x)= 2r_s \left(\phi_0 + \frac{\lVert\bm{x}-\bm{y}\rVert^2}{2} - \log 
    x +\frac{\bm{\alpha} \cdot \bm{x}}{x^2} -\frac{(\bm{\beta}\cdot \bm{x})^2}{x^4}\right)\,,   
 \end{equation} where $\bm{\beta} = \bm{a}/r_E$. From the last term in Eq.~\eqref{eq:time-delay-covariant}, it follows that the second-order correction in the Kerr parameter to the time delay is well-behaved, given that no unphysical singularities are introduced. This discussion shows that the slow spin expansion is well-defined to this order.
 
\begin{figure*}
    \centering
    \includegraphics[width=0.5\linewidth]{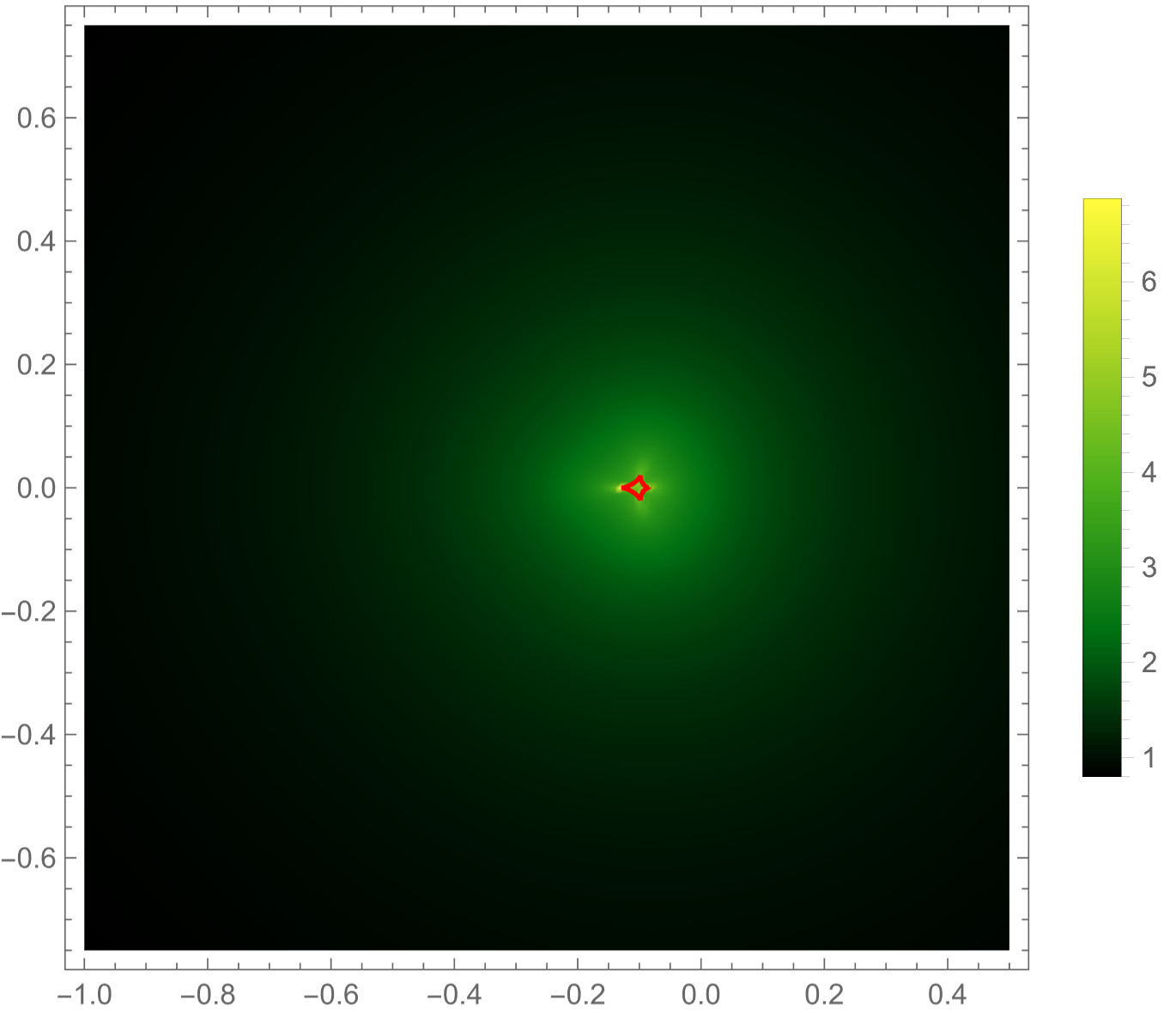}~\includegraphics[width=0.5\linewidth]{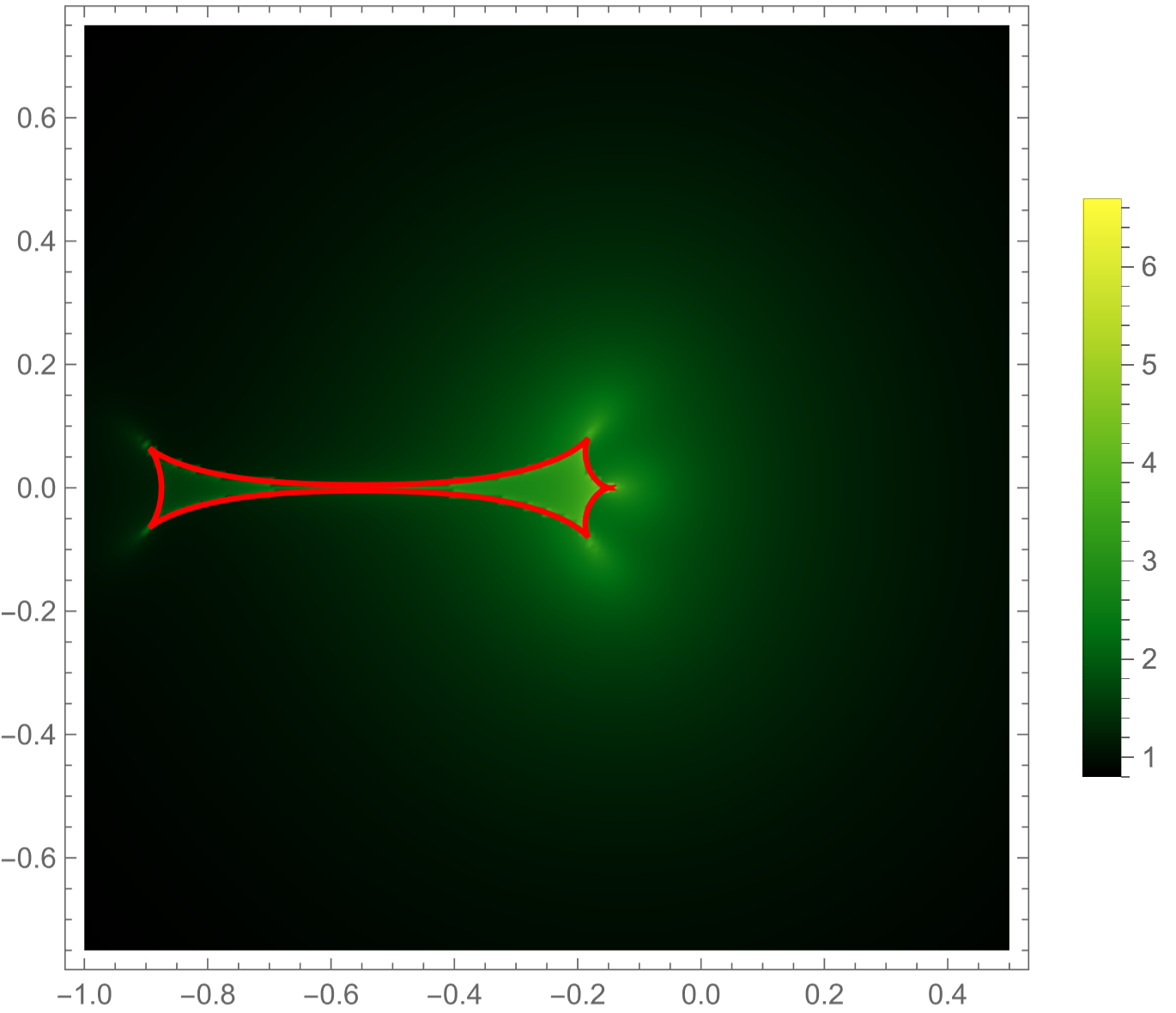}
    \includegraphics[width=0.5\linewidth]{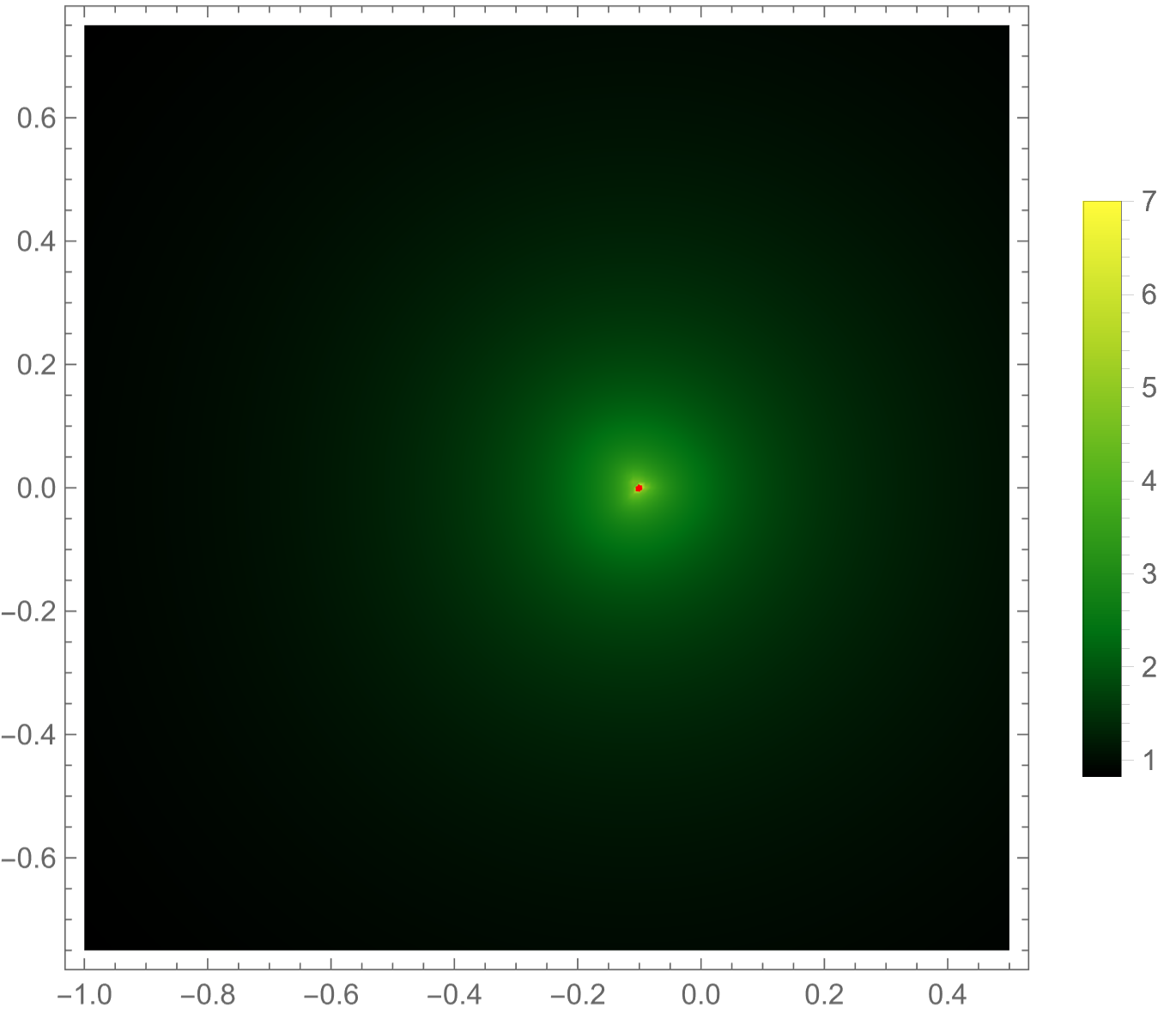}~\includegraphics[width=0.5\linewidth]{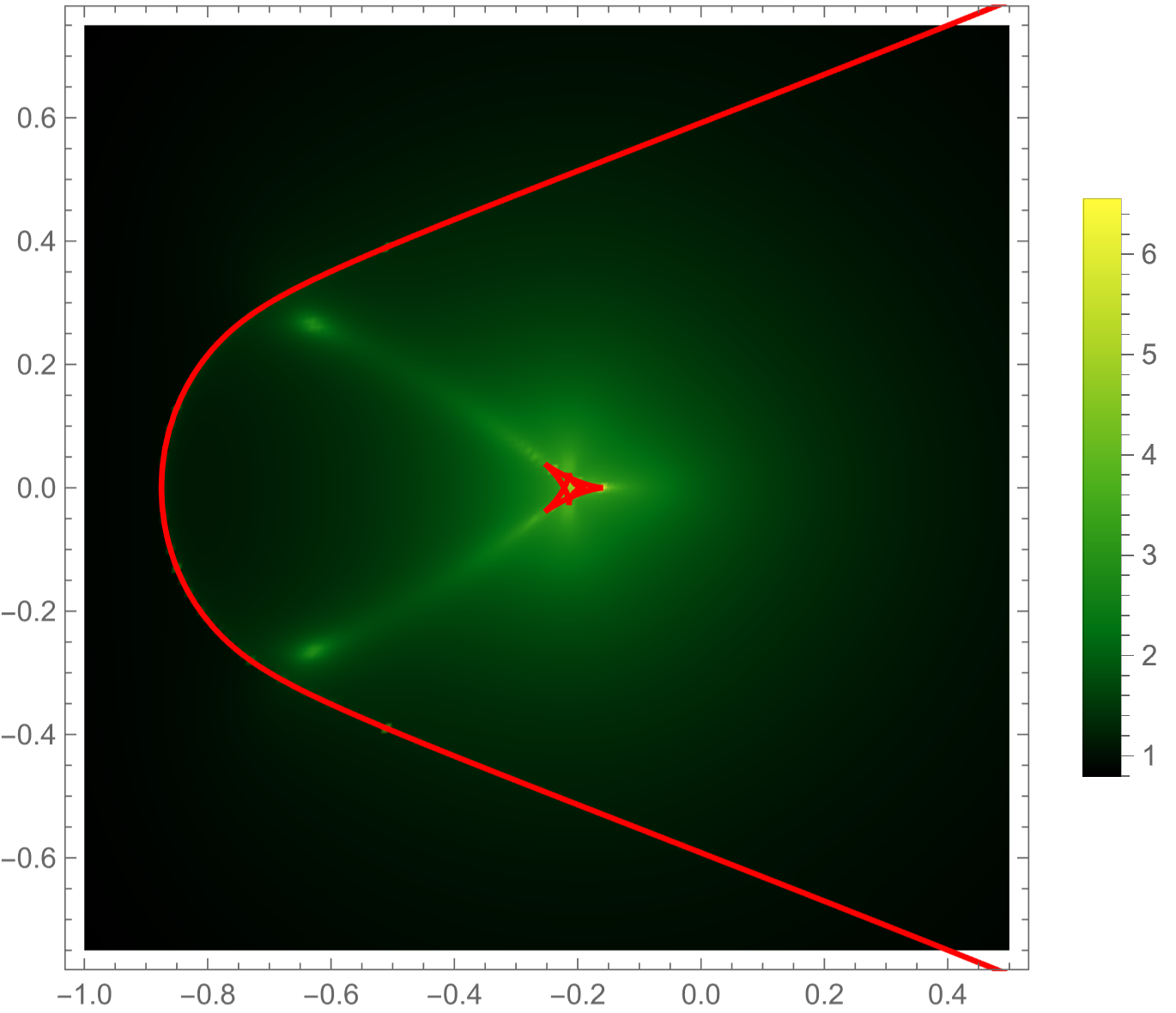}
    \caption{Intensity profile within the geometric optics approximation (Eq.~\eqref{eq:geometric}) for $\alpha=0.1$ (first column) and $\alpha=0.2$ (second column). We used the time delay to first order in the Kerr parameter in Eq.~\eqref{eq:time-delay-dimensionless} (first row) and to second order in $a$ as in Eq.~\eqref{eq:time-delay-covariant} (second row). The caustics are depicted in red.   }
    \label{fig:second-order-alpha}
\end{figure*}

 To estimate the validity of the slow spin approximation, we compare the intensity (computed using the geometric optics approximation Eq.~\eqref{eq:geometric}) obtained with and without the second-order correction in the Kerr parameter to the time delay. The comparison for $\alpha =0.1$  and $\alpha=0.2$ are depicted in the first and second columns of Fig.~\ref{fig:second-order-alpha}. The first row in Fig.~\ref{fig:second-order-alpha} has been computed using Eq.~\eqref{eq:time-delay-dimensionless}, while the second includes the second order correction in $a$ to the time delay in Eq.~\eqref{eq:time-delay-covariant}. The images in the first column (for the normalized spin $\alpha=0.1$) display the same features, and in fact, their intensity profile differs only slightly. For $\alpha = 0.2$ we can already see that the intensity profile computed using the second-order correction in $\alpha$ displays different features to our first-order approximation. Hence, the small limit approximation we have described in this paper is valid for $\alpha\lesssim 0.1$.  Notice that this is not a very stringent restriction for stars as rotating lenses given that astrophysical values of $\alpha$ will be tiny (since $r_E$ is typically very large). 
 
 Finally, a piece of warning is in order: this estimate on the validity of the small spin approximation should be regarded with caution, as it is intended as a rough order of magnitude calculation. We have used the small field approximation of the Kerr metric to compute the next order correction in $a$ to the time delay. Given that the Kerr uniqueness theorem~\cite{PhysRevLett.34.905} is only applicable to perfectly axisymmetric bodies, it is not guaranteed that the exterior of our rotating star is well approximated by the metric in Eq.~\eqref{eq:metric-kerr} (that is to say, Birkhoff's theorem does not generalize to rotating stars).

\end{document}